\documentclass[letterpaper,10pt,conference]{ieeeconf} 
\IEEEoverridecommandlockouts
\usepackage{cite}
\usepackage{amsmath,amssymb,amsfonts}
\usepackage{algorithmic}
\usepackage{graphicx}
\usepackage{textcomp}
\usepackage{xcolor}

\usepackage{booktabs}
\usepackage{mathtools,mathrsfs,optidef}
\mathtoolsset{showonlyrefs=true}

\newtheorem{assumption}{Assumption}
\newtheorem{theorem}{Theorem}

\newtheorem{lemma}{Lemma}

\newtheorem{remark}{Remark}

\newcommand{\C}{\mathbb{C}}
\newcommand{\R}{\mathbb{R}}
\newcommand{\Z}{\mathbb{Z}}
\newcommand{\N}{\mathbb{N}}

\DeclareMathOperator{\argmin}{arg\,min}

\DeclareMathOperator{\rank}{rank}
\DeclareMathOperator{\tr}{tr}

\newcommand{\T}{\top}
\newcommand{\E}{\mathbb{E}}
\newcommand{\Prob}{\mathbb{P}}

\newcommand{\Aset}{\mathcal{A}}
\newcommand{\Bset}{\mathcal{B}}
\newcommand{\Cset}{\mathcal{C}}
\newcommand{\Dset}{\mathcal{D}}
\newcommand{\Iset}{\mathcal{I}}
\newcommand{\Lset}{\mathcal{L}}
\newcommand{\Mset}{\mathcal{M}}
\newcommand{\Oset}{\mathcal{O}}
\newcommand{\Tset}{\mathcal{T}}
\newcommand{\Borel}{\mathscr{B}}
\newcommand{\Fil}{\mathscr{F}}

\newcommand{\midd}{\,\middle|\,}

\usepackage{tikz}
\usetikzlibrary{arrows.meta,calc,positioning,shapes}
\tikzset{
  every node/.style = {outer sep=0.12cm, inner sep=0},
  arrow/.style = {-{Triangle[length=0.25cm]}, thick},
  line/.style = {thick},
  block/.style = {rectangle, draw, minimum height=0.8cm, minimum width=2cm, thick, outer sep=0},
  sum/.style = {thick, circle, draw, inner sep=0, minimum size=6pt, outer sep=0},
  point/.style = {radius=2pt}
}

\usepackage{balance}

\title{\LARGE\textbf{Event-Based Control via Sparsity-Promoting Regularization:\\ A Rollout Approach with Performance Guarantees}}

\author{Shumpei~Nishida and Kunihisa~Okano%
  \thanks{This work was supported by JST SPRING Grant Number JPMJSP2101.}
  \thanks{The authors are with the Graduate School of Science and Engineering,
    Ritsumeikan University,
    Shiga 525-8577 Japan.
    E-mails: {\footnotesize \{\texttt{re0158ff@ed.}, \texttt{kokano@fc.}\}%
    \texttt{ritsumei.ac.jp}}.}
}

\usepackage{fancyhdr}
\usepackage{url}
\fancypagestyle{accfirst}{%
    \fancyhf{}
    \fancyfoot[C]{%
        \fontsize{7}{8}\selectfont
        Published in the Proceedings of the
        2026 American Control Conference (ACC).
    }

}

\begin{document}

\maketitle
\pagestyle{empty}
\thispagestyle{accfirst}

\begin{abstract}
  This paper presents a controller design framework aiming to balance control performance and actuation rate.
  Control performance is evaluated by an infinite-horizon average cost, and the number of control actions is penalized via sparsity-promoting regularization.
  Since the formulated optimal control problem has a combinatorial nature, we employ a rollout algorithm to obtain a tractable suboptimal solution.
  In the proposed scheme, actuation timings are determined through a multistage minimization procedure based on a receding-horizon approach, and the corresponding control inputs are computed online.
  We establish theoretical performance guarantees with respect to periodic control and prove the stability of the closed-loop system.
  The effectiveness of the proposed method is demonstrated through a numerical example.
\end{abstract}

\section{Introduction}
\label{sec:introduction}

This paper investigates a sparse control method that achieves desirable performance under intermittent actuation.
Sparse signals, which remain zero over most of the time horizon, enable networked systems to achieve control objectives with limited resource usage and compact signal representations \cite{nagahara2024survey}.
They can also reduce power consumption by saving actuation durations, which is particularly important in railways \cite{liu2003energy} and electric vehicles \cite{chan2007state}.
These advantages have motivated extensive research on the theoretical foundations of sparse control, including controllability \cite{joseph2020controllability}, stability \cite{sriram2022stabilizability}, and optimal control \cite{lin2013design,nagahara2016maximum}.
Despite these advances, the fundamental challenge lies in reconciling the trade-off between closed-loop performance and input sparsity \cite{jovanovic2016controller}.


Research on controller design under intermittent actuation has followed two main directions.
One line of work studies optimal control problems with constraints on the number of control actions over the entire horizon \cite{imer2006optimal,bommannavar2008optimal,shi2013finite}.
Another approach incorporates sparsity-promoting regularization terms into the optimization problem rather than imposing explicit constraints \cite{cogill2009event,gao2011cardinality}.
These studies provide near-optimal actuation strategies with theoretical performance bounds; however, their problem settings are restricted, such as considering noise-free systems \cite{gao2011cardinality,shi2013finite} or excluding control energy from the objective function \cite{imer2006optimal,bommannavar2008optimal,cogill2009event}.
For a more general setting, \cite{demirel2016trade} has studied packetized dead-beat control with threshold-based event triggering.
While their work quantifies the trade-off between communication cost and control performance, the actuation timings are determined by the predesigned threshold rule rather than being optimized.
This limitation highlights the need for a principled method that jointly optimizes discrete actuation timings and continuous control laws, providing provable performance guarantees.

This paper addresses these limitations by employing the rollout algorithm, a sequential optimization method in the context of dynamic programming \cite{bertsekas2017dynamic-i,bertsekas2012dynamic-ii}.
In a rollout scheme, control laws and scheduling decisions are optimized sequentially over a finite horizon, yielding suboptimal yet effective solutions.
Within this framework, \cite{antunes2014rollout} has addressed quadratic cost minimization under an average communication rate constraint for piecewise-constant control inputs, where the constraint is imposed on the number of input value changes.
Importantly, a theoretical analysis of the controller's performance and stability has been provided.
However, this setting differs from our objective, as we aim to promote sparse signals that remain identically zero over consecutive intervals rather than limiting input variations.
Furthermore, \cite{nishida2024feedback} has studied the simultaneous minimization of a finite-horizon quadratic cost and the number of control actions.
The authors proposed an algorithm that sequentially minimizes the one-step cost with respect to the control input and actuation timing, providing theoretical performance guarantees but no stability guarantees.

This paper develops a rollout-based framework for sparse intermittent actuation, directly addressing the trade-off between control performance and actuation rate.
Using the rollout method, actuation timings are optimized in a receding horizon fashion, and the corresponding control inputs are determined online.
Building on our preliminary work \cite{nishida2024feedback}, which considered only a finite horizon with one-step optimization window and no stability analysis, we establish both performance and stability guarantees for a more general framework.
While our analysis draws inspiration from \cite{antunes2014rollout}, their results cannot be directly applied here, since the trade-off is formulated differently through the sparsity-promoting term.
Finally, a numerical example demonstrates the effectiveness of the proposed method, showing advantages over periodic strategies and $\ell_1$-relaxed approximation of group sparsity.

The remainder of this paper is organized as follows.
In section~\ref{sec:problem_formulation}, we describe the problem formulation.
Section~\ref{sec:main_results} provides some preliminaries and presents the proposed algorithm, followed by a theoretical analysis of our framework in Section~\ref{sec:performance_and_stability_guarantees}.
Section~\ref{sec:simulation_example} illustrates a numerical example, and Section~\ref{sec:conclusion} concludes this paper.

\paragraph*{Notation}
We denote by $\R$ the set of real numbers, by $\N$ the set of natural numbers, and by $\Z_{\geq 0}$ the set of nonnegative integers.
The $n \times n$ identity matrix is represented by $I_n$.
For a positive-semidefinite matrix $P$, its square root is denoted by $P^{1/2}$.
For a set $\Aset$, the indicator function is defined as
\begin{equation}
  \mathbb{I}_\Aset(x) \coloneqq 
  \begin{cases}
    1 & \text{if } x \in \Aset, \\
    0 & \text{otherwise}.
  \end{cases}
\end{equation}
We denote by $\sigma(\cdot)$ the $\sigma$-algebra generated by a collection of sets or random variables.
$\Prob(\cdot)$ denotes probability. 
For a random variable $x$, $\E[x]$ and $\E[x \mid \cdot]$ denote its expectation and conditional expectation, respectively.
For integers $p$ and $q$, the notation $p \equiv 0 \pmod{q}$ indicates that $p$ is an integer multiple of $q$.
\section{Problem Formulation}
\label{sec:problem_formulation}

We consider the following discrete-time linear system:
\begin{align}
  x_{k+1} &= Ax_k + Bu_k + w_k, \label{eq:plant} \\
  y_k &= Cx_k + v_k, \label{eq:measurement}
\end{align}
where $x_k \in \mathbb{R}^{n_x}$, $u_k \in \mathbb{R}^{n_u}$, and $y_k \in \mathbb{R}^{n_y}$ are the state, the control input, and the output, respectively.
The random signals $w_k$ and $v_k$ are assumed to be independent and identically distributed (i.i.d.) Gaussian processes with mean zero and covariance matrices $\Omega_w \succ 0$ and $\Omega_v \succ 0$, respectively.
The initial state $x_0$ follows a Gaussian distribution with mean $\bar{x}_0$ and covariance matrix $\Omega_{x_0} \succeq 0$, and is statistically uncorrelated with both $w_k$ and $v_k$ for all $k \in \Z_{\geq 0}$.
The pairs $(A,B)$ and $(A,\Omega_w^{1/2})$ are assumed to be controllable, and $(A,C)$ is assumed to be observable.

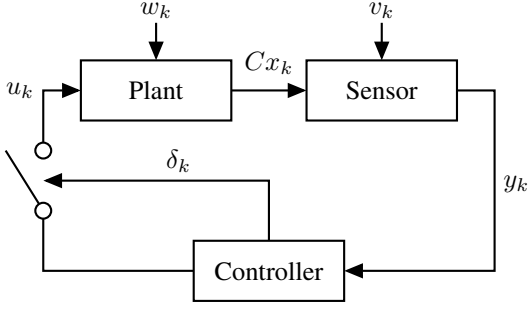
\begin{figure}[t]
  \centering
  \vspace{1em}
  \begin{tikzpicture}
    \node [block] (P){Plant};
    \node [block, right=1cm of P] (S){Sensor};
    \coordinate (mid1) at ($(P)!0.5!(S)$);
    \node [block] (C) at ($(mid1) + (0,-2.4cm)$) {Controller};
    \coordinate (mid2) at ($(mid1)!0.5!(C)$);
    \coordinate (Seast) at ($(S.east) + (0.5cm,0)$);
    \coordinate (Pwest) at ($(P.west) + (-0.5cm,0)$);
    \node [sum] at ($(Pwest)+(0,-0.8cm)$) (sum1){};
    \node [sum] at ($(sum1)+(0,-0.8cm)$) (sum2){};
    \coordinate (w) at ($(P.north)+(0,0.5cm)$);
    \coordinate (v) at ($(S.north)+(0,0.5cm)$);

    \draw [arrow] (w) -- node[above, yshift=2mm] {$w_k$} (P.north);
    \draw [arrow] (v) -- node[above, yshift=2mm] {$v_k$} (S.north);
    \draw [arrow] (P.east) -- node[midway, above=1mm] {$Cx_k$} (S.west);
    \draw [arrow] (S.east) -- (Seast) |- node[pos=0.26, right] {$y_k$} (C.east);
    \draw [line] (C.west) -| (sum2);
    \draw [arrow] (sum1) |- node [midway,left] {$u_k$} (P.west);
    \draw [line] (sum2) -- +(-0.5cm, 0.8cm);
    \draw [arrow] (C.north) |- node[pos=0.7, above] {$\delta_k$} ($(sum1)!0.5!(sum2)$);
    \end{tikzpicture}
  \caption{Considered feedback system.}
  \label{fig:feedback_system}
\end{figure}

The block diagram of the considered system is depicted in Fig.~\ref{fig:feedback_system}.
The controller transmits control signals to the actuators only at selected time instants, while the control input is set to zero at all other time steps.
We represent this actuation strategy by a binary variable $\delta_k \in \{0,1\}$, such that $u_k = 0$ if $\delta_k = 0$, and $u_k \in \R^{n_u}$ otherwise.
At each time $k$, the controller determines $u_k$ and $\delta_k$ based on the information available up to time $k$.
The information set is formally defined as
\begin{align}
  \Iset_k &\coloneqq \{y_i, u_j, \delta_j \colon 0\leq i\leq k,\ 0\leq j\leq k-1 \}, \\
  \Iset_0 &\coloneqq \{y_0\}.
\end{align}
Let $\mu^u_k \colon \Iset_k \to \mathbb{R}^{n_u}$ and $\mu^\delta_k \colon \Iset_k \to \{0,1\}$ denote the control policy and the triggering policy, respectively, both of which are $\sigma(\Iset_k)$-measurable.
Then, the control input $u_k$ and the triggering variable $\delta_k$ are described as
\begin{equation}  \label{eq:control_triggering_policy}
  u_k = \mu^u_k(\Iset_k),\quad \delta_k = \mu^{\delta}_k(\Iset_k).
\end{equation}
We denote the sequences of these policies by $\mu^u \coloneqq \{\mu^u_k\}_{k\in\Z_{\geq 0}}$ and $\mu^{\delta} \coloneqq \{\mu^{\delta}_k\}_{k\in\Z_{\geq 0}}$, respectively.

At each time step, the controller computes a state estimation based on the available information.
The estimation $\hat{x}_k \coloneqq \E[x_k \mid \Iset_k]$\footnote{Throughout this paper, $\E[\cdot\mid\Iset_k]$ denotes shorthand for $\E[\cdot\mid\sigma(\Iset_k)]$.} is updated by
\begin{equation} \label{eq:estimation}
  \hat{x}_{k+1} = A\hat{x}_{k} + Bu_k + G_{k+1}(y_{k+1} - C(A\hat{x}_k + Bu_k))
\end{equation}
with the initial condition $\hat{x}_0 = \bar{x}_0 + G_0(y_0 - C\bar{x}_0)$, where $G_k = \Sigma^-_kC^{\T}(C\Sigma^-_kC^{\T} + \Omega_v)^{-1}$ is the Kalman gain.
The matrix $\Sigma^-_k$ is given by
\begin{align}
  \Sigma^-_k &= A\Sigma_{k-1}A^{\T} + \Omega_w, \label{eq:apripri_error_cov} \\
  \Sigma_k &= \Sigma^-_k - \Sigma^-_kC^{\T}\left(C\Sigma^-_kC^{\T} + \Omega_v\right)^{-1}C\Sigma^-_k, \label{eq:aposteori_error_cov}
\end{align}
where $\Sigma^-_0 = \Omega_{x_0}$ and $\Sigma_k = \E[(x_k - \hat{x}_k)(x_k - \hat{x}_k)^{\T} \mid \Iset_k]$.
By the controllability and the observability assumptions, there exist positive-definite matrices $\tilde{\Sigma}$ and $\Sigma$ such that
\begin{equation} \label{eq:steady_error_cov}
  \lim_{k \to \infty} \Sigma^-_k = \tilde{\Sigma},\quad \lim_{k \to \infty} \Sigma_k = \Sigma,
\end{equation}
and the steady-state Kalman gain is given by
\begin{equation} \label{eq:steady_kalman_gain}
  G = \tilde{\Sigma}C^{\T}\left(C\tilde{\Sigma}C^{\T} + \Omega_v\right)^{-1}.
\end{equation}

The objective of this paper is to design $\mu^u$ and $\mu^{\delta}$ so as to balance the trade-off between control performance and the number of control actions.
The \emph{average control performance} is measured by the following quadratic cost:
\begin{equation} \label{eq:average_control_cost}
  J^a_c(\mu^u,\mu^{\delta}) = \limsup_{N \to \infty} \frac{1}{N} \E\left[ \sum_{k=0}^{N-1} x_k^{\T}Qx_k + u_k^{\T}Ru_k \right],
\end{equation}
where $Q \succeq 0$ and $R \succ 0$.
We assume that the pair $(A,Q^{1/2})$ is observable.
To capture the sparsity of the control input, we introduce the \emph{average actuation rate} as
\begin{equation} \label{eq:average_actuation_times}
  J^a_r(\mu^{\delta}) = \limsup_{N \to \infty} \frac{1}{N} \E\left[ \sum_{k=0}^{N-1} \delta_k \right].
\end{equation}
Using \eqref{eq:average_control_cost} and \eqref{eq:average_actuation_times}, we formulate the co-design problem as
\begin{mini}|s|
  {\mu^u,\mu^{\delta}}
  {J^a(\mu^u,\mu^{\delta}) = J^a_c(\mu^u,\mu^{\delta}) + \theta J^a_r(\mu^{\delta})}
  {\label{prob:codesign_problem_average}}
  {}
  \addConstraint{u_k = 0\quad}{\text{if}\ \delta_k = 0,\quad \forall k \in \Z_{\geq 0},}
\end{mini}
where $\theta > 0$ is a weighting parameter that adjusts the balance between the LQ performance and the actuation rate.
While $\mu^u_k$ takes values in a continuous set, $\mu^{\delta}_k$ takes values in the binary set $\{0,1\}$.
Consequently, Problem~\eqref{prob:codesign_problem_average} is an optimal control problem with mixed discrete and continuous variables, which is generally difficult to solve directly.
To effectively address this problem, we employ the rollout approach as an approximation technique.

\section{Proposed Rollout-Based algorithm}
\label{sec:main_results}


This section presents algorithm design using a rollout approach.
In Section~\ref{subsec:preliminaries}, we provide preliminaries on the rollout method.
In section~\ref{subsec:optimal_periodic_control_policy}, we derive the optimal periodic policy used as a basis for our algorithm, and
Section~\ref{subsec:algorithm_design} presents the proposed algorithm for Problem~\eqref{prob:codesign_problem_average}.

\subsection{Preliminaries for the Rollout Algorithm}
\label{subsec:preliminaries}

The rollout algorithm is a sequential optimization technique in the context of dynamic programming \cite{bertsekas2017dynamic-i,bertsekas2012dynamic-ii}.
In this framework, the optimal value function of the original problem is approximated using a given policy, referred to as the base policy, and a suboptimal policy is obtained through an iterative procedure.

Since directly applying the rollout idea to the average-cost minimization problem~\eqref{prob:codesign_problem_average} is analytically challenging, we instead consider the corresponding infinite-horizon discounted-cost problem, which serves as a tractable surrogate for constructing the rollout-based algorithm.
Formally, we formulate the discounted-cost minimization problem as
\begin{mini}|s|
    {\mu^u,\mu^{\delta}}
    {J^d(\mu^u,\mu^{\delta}) = J^d_c(\mu^u,\mu^{\delta}) + \theta J^d_r(\mu^{\delta})}
    {\label{prob:codesign_problem_discounted}}
    {}
    \addConstraint{u_k = 0\quad}{\text{if}\ \delta_k = 0,\quad \forall k \in \Z_{\geq 0},}
    \end{mini}
where
\begin{align}
    J^d_c(\mu^u,\mu^{\delta}) &= \E\left[ \sum_{k=0}^{\infty} \alpha^k \left(x_k^{\T}Qx_k + u_k^{\T}Ru_k\right) \right], \label{eq:discounted_control_cost}\\
    J^d_r(\mu^{\delta}) &= \E\left[ \sum_{k=0}^{\infty} \alpha^k \delta_k \right], \label{eq:discounted_actuation_times}
    \end{align}
and $\alpha \in (0,1)$ is the discount factor.
By applying the dynamic programming principle, we obtain the following Bellman equation:
\begin{align}
    V^d_k(\Iset_k) = \min_{\substack{u_k,\dots,u_{k+h-1},\\ \delta_k,\dots,\delta_{k+h-1}}}
    &\E\left[ \sum_{i=k}^{k+h-1} \alpha^{i-k} g(x_i, u_i, \delta_i) \right. \notag \\
    &\quad \left. + \alpha^h V^d_{k+h}(\Iset_{k+h}) \,\middle|\, \Iset_k \right], \label{eq:discounted_bellman_equation}
\end{align}
where $V^d_k$ is the optimal value function and $g(x,u,\delta) \coloneqq x^{\T}Qx + u^{\T}Ru + \theta\delta$.

Based on the rollout method, we replace $V^d_{k+h}$ in~\eqref{eq:discounted_bellman_equation} by $\tilde{V}^d_{k+h}$, which is a value function associated with a fixed base policy.
Since $\tilde{V}^d_{k+h}$ is independent of the optimization variables, the problem reduces to the following $h$-step lookahead minimization:
\begin{align}
    \hat{V}^d_k(\Iset_k) = \min_{\substack{u_k,\dots,u_{k+h-1},\\\delta_k,\dots,\delta_{k+h-1}}}
    &\E\left[ \sum_{i=k}^{k+h-1} \alpha^{i-k} g(x_i, u_i, \delta_i) \right. \notag \\
    &\quad \left. + \alpha^h \tilde{V}^d_{k+h}(\Iset_{k+h}) \,\middle|\, \Iset_k \right], \label{eq:discounted_bellman_equation_multistep_lookahead}
\end{align}
At each time $k=\ell h$, the triggering policies over the $h$ steps are obtained by solving \eqref{eq:discounted_bellman_equation_multistep_lookahead} with a fixed control policy in a receding-horizon manner.
This yields a sequence of triggering policies $\mu^{\delta}_k,\dots,\mu^{\delta}_{k+h-1}$, and the corresponding control policies $\mu^u_k,\dots,\mu^u_{k+h-1}$ are then computed online.
The procedure is repeated every $h$ steps.

\subsection{Optimal Periodic Policy and The Corresponding Cost}
\label{subsec:optimal_periodic_control_policy}

In this study, we adopt the periodic policy, in which the control input is applied only at fixed time instants, as the base policy.
Since the objective of this paper is to balance the trade-off between control performance and input sparsity, employing a base policy that yields intermittent actuation provides a consistent choice for approximating the value function.
In what follows, we derive the optimal periodic policy for the discounted-cost minimization problem~\eqref{prob:codesign_problem_discounted} and characterize the corresponding value function.
For analytical purposes, we also derive the optimal periodic policy for Problem~\eqref{prob:codesign_problem_average}.

Let $p \in \N$ denote the control period of the periodic policy.
In this case, the associated triggering policy is given by
\begin{equation} \label{eq:periodic_triggering_policy}
    \mu^\delta_k(\Iset_k) = \begin{cases}
        1 & \text{if } k \equiv 0 \pmod{p}, \\
        0 & \text{otherwise},
        \end{cases}
\end{equation}
meaning that $u_k$ is applied every $p$ steps and $u_k = 0$ otherwise.
Hereafter, we denote the sequences of periodic control and triggering policies with period $p$ by $\mu^{u, \text{per}}$ and $\mu^{\delta, \text{per}}$, respectively.

For a nonnegative integer $\ell \in \Z_{\geq 0}$, we obtain the lifted system describing the state evolution from $k=\ell p$ to $k=(\ell+1)p$ as
\begin{equation} \label{eq:periodic_system}
  x_{(\ell+1)p} = A^p x_{\ell p} + B_p u_{\ell p} + D_p \bar{w}_\ell^{(p)},
\end{equation}
where
\begin{align}
  B_p &\coloneqq A^{p-1}B,\quad D_p \coloneqq \begin{bmatrix}
    A^{p-1} & A^{p-2} & \cdots & I_{n_x}
  \end{bmatrix}, \\
  \bar{w}_\ell^{(p)} &\coloneqq \begin{bmatrix}
    w_{\ell p}^{\T} & w_{\ell p+1}^{\T} & \cdots & w_{(\ell+1)p-1}^{\T}
  \end{bmatrix}^\T.
\end{align}
The random vector $\bar{w}_\ell^{(p)}$ follows an independent Gaussian distribution with mean zero, whose covariance matrix $\Omega_w^{(p)}$ is given by
\begin{equation}
  \Omega_w^{(p)} = I_p \otimes \Omega_w.
\end{equation}
To reformulate the discounted-cost problem~\eqref{prob:codesign_problem_discounted} and the average-cost problem~\eqref{prob:codesign_problem_average} based on the lifted representation~\eqref{eq:periodic_system}, we introduce the matrices $Q_p$, $S_p$, and $R_p$ as
\begin{align}
  Q_p &\coloneqq \sum_{i=0}^{p-1} (A^i)^{\T} Q A^i,\quad S_p \coloneqq \sum_{i=1}^{p-1} (A^i)^{\T} Q A^{i-1} B, \\
  R_p &\coloneqq R + \sum_{i=1}^{p-1} B^{\T}(A^{i-1})^{\T} Q A^{i-1} B.
\end{align}
We also define the function $c_p \colon \R^{n_x} \times \R^{n_u} \to \R$ as $c_p(x,u) \coloneqq x^{\T} Q_p x + 2x^{\T} S_p u + u^{\T} R_p u$.
Then, using \eqref{eq:periodic_system}, $J^d_c$ and $J^a_c$ can be expressed as
\begin{align}
  J^d_c(\mu^u, \mu^{\delta, \text{per}}) &= \E\left[\sum_{\ell=0}^{\infty} \alpha^{\ell p} c_p(x_{\ell p}, u_{\ell p})\right] + d^d(p), \label{eq:control_cost_periodic_discounted} \\
  J^a_c(\mu^u, \mu^{\delta, \text{per}}) &= \limsup_{K \to \infty} \frac{1}{pK} \E\left[\sum_{\ell=0}^{K-1} c_p(x_{\ell p}, u_{\ell p})\right] \notag \\
  &\quad + \frac{1}{p} d^a(p), \label{eq:control_cost_periodic_average}
\end{align}
where $d^d(p)$ and $d^a(p)$ are given by
\begin{align}
  d^d(p) &\coloneqq \frac{1}{1-\alpha^p} \sum_{i=1}^{p-1} \alpha^i \sum_{j=0}^{i-1} \tr\left(Q A^j \Omega_w (A^j)^\T\right), \\
  d^a(p) &\coloneqq \sum_{i=1}^{p-1} \sum_{j=0}^{i-1} \tr\left(Q A^j \Omega_w (A^j)^\T\right).
\end{align}
Since $\mu^{\delta, \text{per}}$ is fixed independently of $\mu^u$, the terms $J^d_r$ and $J^a_r$ can be computed explicitly as
\begin{equation}
  J^d_r(\mu^{\delta, \text{per}}) = \frac{1}{1-\alpha^p},\quad J^a_r(\mu^{\delta, \text{per}}) = \frac{1}{p}.
\end{equation}
Therefore, once $\mu^{\delta,\text{per}}$ is given as the triggering policy, Problem~\eqref{prob:codesign_problem_discounted} reduces to finding $\mu^u$ that minimizes $J^d_c$.
Similarly, for the average-cost problem~\eqref{prob:codesign_problem_average}, it suffices to find $\mu^u$ that minimizes $J^a_c$.

Suppose the following assumption holds.
\begin{assumption}
  \label{assump:eigenvalue_condition}
  Let $\lambda_a$ and $\lambda_b$ be any two distinct eigenvalues of $A$.
  Then it holds that
  \begin{equation} \label{eq:eigenvalue_condition}
    \lambda_a \neq \lambda_b \exp\left( \frac{2\pi q \sqrt{-1}}{p} \right), \quad \forall q \in \mathbb{Z}.
  \end{equation}
\end{assumption}
This assumption ensures that $(A^p, B_p)$ is stabilizable when $(A,B)$ is controllable \cite{pasand2018controllability}.
Then, the control policy that minimizes $J^d_c$ under control period $p$ can be obtained using standard optimal control techniques \cite{bertsekas1976dynamic}:
\begin{equation} \label{eq:optimal_periodic_control_discounted}
  \mu^{u,\text{per}}_k = \begin{cases}
    \tilde{F}_p \hat{x}_k & \text{if } k \equiv 0 \pmod{p}, \\
    0 & \text{otherwise},
  \end{cases}
\end{equation}
where
\begin{equation} \label{eq:feedback_gain_discounted}
  \tilde{F}_p = - \left(\alpha^p B_p^{\T}\tilde{P}_pB_p + R_p\right)^{-1}\left(\alpha^p B_p^{\T}\tilde{P}_pA^p + S_p^{\T}\right),
\end{equation}
and $\tilde{P}_p$ is the positive-definite solution of the following algebraic Riccati equation\footnote{If $(A,Q^{1/2})$ is observable, then $(A^p,Q_p^{1/2})$ is also observable, which ensures the existence of a positive-definite solution (see Appendix~\ref{append:preliminary_lemmas}).}:
\begin{align}
  \tilde{P}_p &= Q_p + \alpha^p (A^p)^{\T}\tilde{P}_p A^p - M_p, \label{eq:algebraic_riccati_discounted} \\
  M_p &\coloneqq \tilde{F}_p^{\T} \left(\alpha^p B_p^{\T}\tilde{P}_pB_p + R_p\right) \tilde{F}_p.
\end{align}
The discounted cost associated with the periodic policy $(\mu^{u, \text{per}}, \mu^{\delta, \text{per}})$ is given by
\begin{align}
  &J^d(\mu^{u, \text{per}}, \mu^{\delta, \text{per}}) = \E\left[x_0^{\T}\tilde{P}_p x_0 \,\middle|\, \Iset_0\right] \notag \\
  &\quad + \frac{\alpha^p}{1 - \alpha^p} \tr\left(\tilde{P}_p D_p \Omega_w^{(p)} D_p^{\T}\right) + \sum_{\ell=0}^{\infty} \alpha^{\ell p} \tr\left(M_p \Sigma_{\ell p}\right) \notag \\
  &\quad + d^d(p) + \frac{\theta}{1 - \alpha^p}. \label{eq:optimal_periodic_cost_discounted}
\end{align}
Furthermore, the periodic control law that minimizes $J^a$ under control period $p$ is the same as \eqref{eq:optimal_periodic_control_discounted}, with $\alpha=1$ substituted into \eqref{eq:feedback_gain_discounted} and \eqref{eq:algebraic_riccati_discounted}, and the associated average cost is calculated as follows:
\begin{align}
  J^a(\mu^{u, \text{per}}, \mu^{\delta, \text{per}}) &= \frac{1}{p} \Big[\tr\left(\tilde{P}_p D_p \Omega_w^{(p)} D_p^{\T}\right) + \tr\left(M_p \Sigma\right) \\
  &\quad + d^a(p)\Big] + \frac{\theta}{p}. \label{eq:optimal_periodic_cost_average}
\end{align}

In the following section, we investigate the rollout-based algorithm for Problem~\eqref{prob:codesign_problem_discounted} using the periodic policy $(\mu^{u, \text{per}}, \mu^{\delta, \text{per}})$ as the base policy, and relate the results to Problem~\eqref{prob:codesign_problem_average}.


\subsection{Algorithm Design}
\label{subsec:algorithm_design}

Before proceeding to the algorithm design, we introduce some notation.
Let $\Tset$ denote the set of sequences of triggering variables in which the first $h$ elements of $\{\delta_k\}_{k \in \Z_{\geq 0}}$ are free parameters, while the remaining elements follow the periodic triggering policy $\mu^{\delta, \text{per}}$.
Since $\Tset$ contains $2^h$ possible sequences, there exists a bijective correspondence between $\Tset$ and the index set $\Mset \coloneqq \{1,2,\dots,2^h\}$.
Hence, for each $m \in \Mset$, an arbitrary sequence in $\Tset$ can be represented as $\{\delta^{(m)}_k\}_{k \in \Z_{\geq 0}}$.
With this notation, we characterize the triggering sequence $\{\delta^{(m)}_k\}_{k \in \Z_{\geq 0}} \in \Tset,\ m \in \Mset$ as
\begin{align}
  \delta^{(m)}_k &= \rho^{(m)}_k, \quad k \in \{0,1,\dots,h-1\}, \\
  \delta^{(m)}_k &= 
  \begin{cases}
    1 & \text{if } k \equiv 0 \pmod{p}, \\
    0 & \text{otherwise},
  \end{cases}
  \quad k \geq h,
\end{align}
where $\rho^{(m)}_k \in \{0,1\}$ for all $k \in \Z_{\geq 0}$.
In particular, for $m=1$, we define
\begin{equation}
  \rho^{(1)}_k = 
  \begin{cases}
    1 & \text{if } k \equiv 0 \pmod{p}, \\
    0 & \text{otherwise},
  \end{cases}
  \quad k \in \{0,1,\dots,h-1\}.
\end{equation}

We are now ready to construct the algorithm for Problem~\eqref{prob:codesign_problem_discounted} with the base policy $(\mu^{u, \text{per}}, \mu^{\delta, \text{per}})$.
Let $h$ be chosen so that $h/p \in \N$.
From \eqref{eq:optimal_periodic_cost_discounted}, the value function under the periodic policy $(\mu^{u, \text{per}}, \mu^{\delta, \text{per}})$ for the discounted-cost problem~\eqref{prob:codesign_problem_discounted} is obtained as
\begin{align}
  &V^{d, \text{per}}_{\ell p}(\Iset_{\ell p}) = \E\left[x_{\ell p}^{\T}\tilde{P}_p x_{\ell p} \,\middle|\, \Iset_{\ell p}\right] \notag \\
  &\quad + \frac{\alpha^p}{1 - \alpha^p} \tr\left(\tilde{P}_p D_p \Omega_w^{(p)} D_p^{\T}\right) + \sum_{j=\ell}^{\infty} \alpha^{(j-\ell)p} \tr\left(M_p \Sigma_{jp}\right) \notag \\
  &\quad + d^d(p) + \frac{\theta}{1 - \alpha^p}, \label{eq:value_function_discounted}
\end{align}
for $k=\ell p$.
Since $h/p \in \N$, we have $k+h \equiv 0 \pmod{p}$ when $k=\ell p$.
Substituting $V^{d,\text{per}}_{k+h}$ for $V^d_{k+h}$ in \eqref{eq:discounted_bellman_equation_multistep_lookahead} and rewriting $V^d_k$ as $\hat{V}^d_k$, we obtain the following equation:
\begin{align}
  &\hat{V}^d_k(\Iset_k) = \min_{\substack{u_k,\dots,u_{k+h-1},\\ \delta_k,\dots,\delta_{k+h-1}}} 
  \E\left[ \sum_{i=k}^{k+h-1} \alpha^{i-k} g(x_i,u_i,\delta_i) \right. \notag \\
  &\quad \left. + \alpha^h \E\left[x_{k+h}^{\T}\tilde{P}_p x_{k+h} \,\middle|\, \Iset_{k+h}\right] \,\middle|\, \Iset_k \right] + \alpha^h \eta^d_k, \label{eq:discounted_bellman_equation_multistep_lookahead_expanded}
\end{align}
where
\begin{align}
  &\eta^d_k \coloneqq \frac{\alpha^p}{1 - \alpha^p} \tr\left(\tilde{P}_p D_p \Omega_w^{(p)} D_p^{\T}\right) \notag \\
  & + \sum_{j=(k+h)/p}^{\infty} \alpha^{(j-(k+h)/p)p} \tr\left(M_p \Sigma_{jp}\right) + d^d(p) + \frac{\theta}{1 - \alpha^p}.
\end{align}
By the tower property of conditional expectation \cite[Th.~5.1.6]{durrett2010probability}, we have
\begin{equation} \label{eq:tower_property}
  \E\left[ \E\left[x_{k+h}^{\T}\tilde{P}_p x_{k+h} \,\middle|\, \Iset_{k+h}\right] \,\middle|\, \Iset_k \right] = \E\left[x_{k+h}^{\T}\tilde{P}_p x_{k+h} \,\middle|\, \Iset_k\right].
\end{equation}  
Since $\eta^d_k$ does not depend on $\mu^u$ and $\mu^{\delta}$, it suffices to minimize the following cost to obtain the policy minimizing $\hat{V}^d_k$ in \eqref{eq:discounted_bellman_equation_multistep_lookahead_expanded}:
\begin{equation} \label{eq:cost_to_minimize}
  \E\left[ \sum_{i=k}^{k+h-1} \alpha^{i-k} g(x_i,u_i,\delta_i) + \alpha^h x_{k+h}^{\T}\tilde{P}_p x_{k+h} \,\middle|\, \Iset_k \right].
\end{equation}
Note that if $m \in \Mset$ is fixed, this cost \eqref{eq:cost_to_minimize} is a standard finite-horizon discounted cost conditioned on $\Iset_k$.
Therefore, for a fixed $m \in \Mset$, we can derive the control input minimizing \eqref{eq:cost_to_minimize} as
\begin{equation} \label{eq:rollout_control_input}
  u_k = \rho^{(m)}_s F^{(m)}_s \hat{x}_k, \quad s \in \{0,1,\dots,h-1\},
\end{equation}
where for any $s \in \{0,1,\dots,h-1\}$ and $m \in \Mset$,
\begin{equation}
  F^{(m)}_s = -\alpha \big(\alpha B^{\T}P^{(m)}_{s+1}B + R\big)^{-1} B^{\T}P^{(m)}_{s+1}A,
\end{equation}
and $P^{(m)}_s$ is given recursively by
\begin{align}
  &P^{(m)}_h = \tilde{P}_p, \\
  &P^{(m)}_s = Q + \alpha A^{\T}P^{(m)}_{s+1}A \notag \\
  &- \alpha^2 \rho^{(m)}_s A^{\T}P^{(m)}_{s+1}B \left(\alpha B^{\T}P^{(m)}_{s+1}B + R\right)^{-1} B^{\T}P^{(m)}_{s+1}A, \notag \\
  &\qquad s \in \{0,1,\dots,h-1\}. \label{eq:riccati_equation_multistep_lookahead}
\end{align}
Hence, for a given $m \in \Mset$, $\hat{V}^d_k$ can be expressed as
\begin{equation} \label{eq:rollout_cost}
  \hat{V}^d_k(\Iset_k) = \hat{x}_k^{\T} P^{(m)}_0 \hat{x}_k + \tr\left(P^{(m)}_0 \Sigma_k\right) + \beta^{(m)}_k + \gamma^{(m)},
\end{equation}
where for each $m \in \Mset$,
\begin{align}
  \beta^{(m)}_k &\coloneqq \sum_{\tau=0}^{h-1} \alpha^{\tau}\left( \tr\left(P^{(m)}_{\tau+1}\Omega_w\right) + \tr\left(M^{(m)}_\tau \Sigma_{k+\tau}\right)\right), \\
  M^{(m)}_\tau &\coloneqq (F^{(m)}_\tau)^{\T}\left(\alpha B^{\T}P^{(m)}_{\tau+1}B + R\right)F^{(m)}_\tau, \\
  \gamma^{(m)} &\coloneqq \theta \sum_{\tau=0}^{h-1} \alpha^{\tau}\rho^{(m)}_\tau.
\end{align}
Therefore, by selecting $m \in \Mset$ that minimizes $\hat{V}^d_k$ in \eqref{eq:rollout_cost} and computing the corresponding control input using \eqref{eq:rollout_control_input}, we can construct the policies over the interval $[k,k+h-1]$.

The preceding discussion has focused on the discounted-cost problem~\eqref{prob:codesign_problem_discounted} with $\alpha \in (0,1)$.
We now construct the algorithm for the average-cost problem~\eqref{prob:codesign_problem_average} by taking the limit as $\alpha \uparrow 1$ in \eqref{eq:discounted_bellman_equation_multistep_lookahead_expanded}.
Although $\eta^d_k$ diverges in this limit, it does not affect the optimization procedure since it is independent of the optimization variables.
Furthermore, since $(A^p, B_p)$ is stabilizable, $(A^p, Q_p^{1/2})$ is observable, and $R_p$ is positive definite, a stabilizing solution $\tilde{P}_p$ exists at $\alpha=1$, and $\tilde{P}_p$ is continuous on $\alpha \in (0,1]$.
Consequently, $P^{(m)}_k$, $F^{(m)}_k$, and $M^{(m)}_k$ are also continuous on $\alpha \in (0,1]$, and hence \eqref{eq:rollout_control_input} and \eqref{eq:rollout_cost} remain valid at $\alpha=1$.

Summarizing the above discussion, the proposed algorithm for Problem~\eqref{prob:codesign_problem_average} is presented as follows.

\paragraph*{Algorithm~1}
\begin{enumerate}
\renewcommand{\labelenumi}{(\roman{enumi})}
  \item At time $k = \ell h,\ \ell \in \Z_{\geq 0}$, compute
  \begin{align} \label{eq:rollout_triggering_policy}
    \varphi(\Iset_{\ell h}) \coloneqq \underset{m \in \Mset}{\argmin}\ &\hat{x}_{\ell h}^{\T}P^{(m)}_0\hat{x}_{\ell h} + \tr\left(P^{(m)}_0\Sigma_{\ell h}\right) \notag \\
    &\quad + \beta^{(m)}_{\ell h} + \gamma^{(m)}.
  \end{align}
  \item Determine the triggering variables over the interval $[\ell h,(\ell+1)h-1]$ by
  \begin{equation}
    \delta_k = \rho^{(\varphi(\Iset_{\ell h}))}_{k - \ell h},
  \end{equation}
  and compute the corresponding control input as
  \begin{equation} \label{eq:rollout_control_policy}
    u_k = \rho^{(\varphi(\Iset_{\ell h}))}_{k - \ell h} F^{(\varphi(\Iset_{\ell h}))}_{k - \ell h} \hat{x}_k.
  \end{equation}
  \item Repeat steps (i) and (ii) every $h$ steps.
\end{enumerate}
Therefore, the control policy $\mu^u_k$ and the triggering policy $\mu^{\delta}_k$ are determined by Algorithm~1 as
\begin{align}
  (\mu^u_k(\Iset_k), \mu^{\delta}_k(\Iset_k)) 
  &= \left(\rho^{(\varphi(\Iset_{\ell h}))}_{k - \ell h} F^{(\varphi(\Iset_{\ell h}))}_{k - \ell h} \hat{x}_k, \rho^{(\varphi(\Iset_{\ell h}))}_{k - \ell h}\right), \\
  &\ell h \leq k < (\ell+1)h,\quad \ell \in \Z_{\geq 0}. \label{eq:proposed_policy_design}
\end{align}


\section{Performance and Stability Guarantees of the Proposed Algorithm}
\label{sec:performance_and_stability_guarantees}

This section establishes theoretical performance guarantees for Algorithm~1 in Problem~\eqref{prob:codesign_problem_average}, in comparison with periodic control.
We also provide a stability result for the closed-loop system under the policy obtained by Algorithm~1.

We impose the following assumptions for the analysis.

\begin{assumption}
  \label{assump:performance_and_stability_guarantees}
  \mbox{}
  \begin{enumerate}
    \renewcommand{\labelenumi}{(\roman{enumi})}
    \item $\rank(C) = \rank(C^{\T}) = n_x$.
    \item $Q \succ 0$.
    \item $\Omega_{x_0}$ is a solution of the following Riccati equation:
    \begin{align}
      \Omega_{x_0} &= \Omega_w + A\Omega_{x_0}A^{\T} \\
      & - A\Omega_{x_0}C^{\T} \left(C\Omega_{x_0}C^{\T} + \Omega_v\right)^{-1} C\Omega_{x_0}A^{\T}. \label{eq:riccati_X0}
    \end{align}
  \end{enumerate}
\end{assumption}

Assumption~\ref{assump:performance_and_stability_guarantees}(i) ensures that the steady-state Kalman gain $G$ has full row rank, since the covariance matrix $\tilde{\Sigma}$ is positive definite.
Assumption~\ref{assump:performance_and_stability_guarantees}(ii) guarantees that the minimum eigenvalue of $Q$ is strictly positive \cite[Lem.~10.4.1]{bernstein2018scalar}.
When Assumption~\ref{assump:performance_and_stability_guarantees}(iii) holds, $\Omega_{x_0}$ is positive definite since $(A,\Omega_w^{1/2})$ is controllable.
Furthermore, this assumption implies $G_k = G$ for all $k$, i.e., the Kalman filter is stationary, which is the case of interest in this study.
In the following, we write $\beta^{(m)}$ instead of $\beta^{(m)}_k$, since $\beta^{(m)}_k$ is independent of the time index $k$ under Assumption~\ref{assump:performance_and_stability_guarantees}.


Let $(\mu^{u, \text{ro}}, \mu^{\delta, \text{ro}})$ denote the sequence of the policies obtained by Algorithm~1.
The following theorem provides a theoretical performance bound.

\begin{theorem}
  \label{thm:performance_guarantee}
  Suppose that Assumptions~\ref{assump:eigenvalue_condition} and \ref{assump:performance_and_stability_guarantees} hold.
  Then, the following inequality holds:
  \begin{equation} \label{eq:performance_guarantee}
    J^a(\mu^{u, \text{ro}}, \mu^{\delta, \text{ro}}) 
    \leq J^a(\mu^{u, \text{per}}, \mu^{\delta, \text{per}}) + \frac{1}{h}.
  \end{equation}
\end{theorem}
\begin{proof}
  See Appendix~\ref{append:proof_of_performance_guarantee}.
\end{proof}

The additive term $1/h$ in \eqref{eq:performance_guarantee} appears due to technical reasons.
When $h$ is sufficiently large, this term becomes negligible.
In such cases, Algorithm~1 is executed only at sparse time instants, with most actuation timings and feedback gains being computed in an open-loop fashion.
Consequently, the control policy $\mu^{u, \text{ro}}_k$ and the triggering policy $\mu^{\delta, \text{ro}}_k$ are determined primarily by statistical information, such as $\Omega_w$, $\Omega_v$, and $\Sigma$, rather than by actual disturbances.
This resembles the optimal periodic control, where the periodic policy is determined from stationary statistics via minimizing \eqref{eq:optimal_periodic_cost_average}.
In such situations, \eqref{eq:performance_guarantee} shows that Algorithm~1 guarantees performance no worse than the optimal periodic policy.

\begin{remark}
  The horizon length $h$ affects both the computational burden and the achieved performance.
  In particular, the set $\Mset$ has cardinality $2^h$, resulting in exponential growth of the algorithmic complexity with $h$.
  Meanwhile, Algorithm~1 is executed only every $h$ steps, and thus increasing $h$ reduces the execution frequency.
  Therefore, selecting $h$ involves a trade-off between the computational complexity and the execution frequency, which is not fully addressed in this paper.
  Deriving a principled rule for selecting $h$ is left for future work.
\end{remark}

For a given $m \in \Mset$, we define $V_m \colon \R^{n_x} \to [0,\infty)$ by
\begin{equation} \label{eq:foster_lyapunov_function}
  V_m(x) \coloneqq x^{\T} P_0^{(m)} x,\quad \forall x \in \R^{n_x}.
\end{equation}
For notational convenience, we introduce $\chi_\ell \coloneqq x_{\ell h}$ and $\hat{\chi}_\ell \coloneqq \hat{x}_{\ell h}$ for $\ell \in \Z_{\geq 0}$.
The following lemmas are used to establish Theorem~\ref{thm:performance_guarantee}.

\begin{lemma}
  \label{lemma:markov_chain}
  Under Assumptions~\ref{assump:eigenvalue_condition} and \ref{assump:performance_and_stability_guarantees}, 
  the stochastic process $\{\hat{\chi}_\ell\}_{\ell \in \Z_{\geq 0}}$ induced by $(\mu^{u, \text{ro}}, \mu^{\delta, \text{ro}})$ is a Markov chain.
\end{lemma}
\begin{proof}
  See Appendix~\ref{append:markov_chain}.
\end{proof}

\begin{lemma} \label{lemma:foster_lyapunov}
  If Assumptions~\ref{assump:eigenvalue_condition} and \ref{assump:performance_and_stability_guarantees} hold, then it holds that
  \begin{equation} \label{eq:foster_lyapunov}
    \limsup_{L \to \infty} \E\left[V_1(\chi_L) \midd \hat{\chi}_0\right] < \infty.
  \end{equation}
\end{lemma}
\begin{proof}
  See Appendix~\ref{append:foster_lyapunov}.
\end{proof}

Since Algorithm~1 determines $\varphi(\Iset_{\ell h})$ by minimizing a criterion that depends on $\Iset_{\ell h}$ only through $\hat\chi_\ell$, we can regard $\varphi$ as a Borel measurable function of $\hat\chi_\ell$ and write $\varphi(\Iset_{\ell h}) = \varphi(\hat\chi_\ell)$.
Then, we define $f \colon \R^{n_x}\to[0,\infty)$ as
\begin{align}
  f(x) &\coloneqq x^{\T} \left(P_0^{(1)} - P_0^{(\varphi(x))}\right) x 
  + \tr\left(\left(P_0^{(1)} - P_0^{(\varphi(x))}\right) \Sigma\right) \notag \\ 
  &\quad + \beta^{(1)} - \beta^{(\varphi(x))} + \gamma^{(1)} - \gamma^{(\varphi(x))}.
  \label{eq:def_fm}
\end{align}
By construction of Algorithm~1, we have $f(x) \geq 0$ for all $x \in \R^{n_x}$.
The following theorem states that the Markov chain $\{\hat{\chi}_\ell\}_{\ell \in \Z_{\geq 0}}$ is ergodic, i.e., positive Harris recurrent \cite[Ch.~14]{meyn2012markov}.

\begin{theorem}
  \label{thm:ergodicity}
  Define the mapping $\tilde{f} \coloneqq f + 1$.
  Let Assumptions~\ref{assump:eigenvalue_condition} and \ref{assump:performance_and_stability_guarantees} hold.
  Then, under the policy $(\mu^{u, \text{ro}}, \mu^{\delta, \text{ro}})$, the Markov chain $\{\hat{\chi}_\ell\}_{\ell \in \Z_{\geq 0}}$ is $\tilde{f}$-ergodic.
\end{theorem}
\begin{proof}
  See Appendix~\ref{append:proof_of_ergodicity}.
\end{proof}

Furthermore, Lemma~\ref{lemma:foster_lyapunov} implies that Algorithm~1 ensures mean-square stability of the closed-loop system, as stated below.

\begin{theorem}
  \label{thm:stability_guarantee}
  Suppose that Assumptions~\ref{assump:eigenvalue_condition} and \ref{assump:performance_and_stability_guarantees} hold.
  Then, the closed-loop system under $(\mu^{u, \text{ro}}, \mu^{\delta, \text{ro}})$ is mean-square stable, i.e.,
  \begin{equation} \label{eq:mean_square_stability}
    \sup_{k \in \Z_{\geq 0}} \E\left[\|x_k\|^2\right] < \infty.
  \end{equation}
\end{theorem}
\begin{proof}
  See Appendix~\ref{append:proof_of_stability_guarantee}.
\end{proof}

\section{Illustrative Example}
\label{sec:simulation_example}

Consider a system consisting of two masses $m_1$ and $m_2$ connected by a spring with spring constant $\kappa_s$ \cite{antunes2014rollout}.
We set $m_1 = m_2 = 1$ and $\kappa_s = 2\pi^2$.
The control input and the process noise affect only mass $m_1$.
The model of this system is described by the following stochastic differential equations:
\begin{align}
  dx(t) &= \left(A_C x(t) + B_C u(t)\right) dt + D_C dw(t), \label{eq:plant_simu} \\
  dy(t) &= C_C x(t) dt + dv(t), \label{eq:measurement_simu}
\end{align}
where
\begin{align}
  A_C &= \begin{bmatrix}
    0 & 0 & 1 & 0\\
    0 & 0 & 0 & 1\\
    -\frac{\kappa_s}{m_1} & \frac{\kappa_s}{m_1} & 0 & 0\\
    \frac{\kappa_s}{m_2} & -\frac{\kappa_s}{m_2} & 0 & 0
  \end{bmatrix},\quad
  B_C = \begin{bmatrix}
    0\\
    0\\
    \frac{1}{m_1}\\
    0
  \end{bmatrix}, \\
  D_C &= \begin{bmatrix}
    0\\
    0\\
    0.4\\
    0
  \end{bmatrix},\quad
  C_C = \begin{bmatrix}
    1 & 0 & 0 & 0\\
    0 & 1 & 0 & 0
  \end{bmatrix}, \label{eq:DC_CC}
\end{align}
and $x(t) = [x^1(t)\ x^2(t)\ \dot{x}^1(t)\ \dot{x}^2(t)]^{\T} \in \R^4$ denotes the state vector, where $x^1(t)$ and $x^2(t)$ are the displacements of masses $m_1$ and $m_2$, respectively.
We assume that $w(t)$ and $v(t)$ are independent Wiener processes with zero mean, whose incremental covariances are $I_{n_x}dt$ and $10^{-5} I_{n_y}dt$, respectively.
By discretizing \eqref{eq:plant_simu} and \eqref{eq:measurement_simu} with the sampling period $t_s = 0.1$, we obtain the discrete-time system \eqref{eq:plant} and \eqref{eq:measurement}.
The discretized disturbances $w_k$ and $v_k$ are i.i.d. Gaussian random vectors with zero mean and covariance matrices
\begin{equation}
  \Omega_w = \int_0^{t_s} e^{A_C \tau} D_C D_C^{\T} e^{A_C^{\T} \tau}\, d\tau, \quad 
  \Omega_v = \frac{10^{-5}}{t_s} I_{n_y}.
\end{equation}
For the discretized system, we choose $\bar{x}_0 = [1\ -1\ 0\ 0]^{\T}$ and let $\Omega_{x_0}$ satisfy \eqref{eq:riccati_X0}.
We set the control horizon to 600, and $Q$ and $R$ are given by
\begin{align}
  Q &= \begin{bmatrix}
    0.1336 & -0.0936 & -0.0327 & 0.0347\\
    -0.0936 & 0.1336 & 0.0347 & -0.0327\\
    -0.0327 & 0.0347 & 0.0377 & 0.0024\\
    0.0347 & -0.0327 & 0.0024 & 0.0377
  \end{bmatrix},\\
  R &= 0.1.
\end{align}

\begin{remark}
  The matrix $C_c$ in \eqref{eq:DC_CC} does not meet Assumption~\ref{assump:performance_and_stability_guarantees}(i).
  Therefore, the theoretical guarantees established under Assumption~\ref{assump:performance_and_stability_guarantees} are not claimed to apply to this example.
  Nonetheless, the following simulation results provide an empirical assessment beyond the sufficient conditions used in the analysis, suggesting that the proposed method can still improve performance under limited control actions even when the theoretical conditions are not satisfied.
\end{remark}

We evaluate three methods for Problem~\eqref{prob:codesign_problem_average}.
The first one is Algorithm~1 presented in Section~\ref{subsec:preliminaries}, with $h=6$.
At $k=6\ell$ for $\ell \in \Z_{\geq 0}$, the triggering variables and the corresponding control inputs are computed according to \eqref{eq:proposed_policy_design}.
The second method is the periodic control described in Section~\ref{subsec:preliminaries}, with $p \in \{1,2,3,6\}$.
The third method is an $\ell_1$-relaxation approach, where $\delta_k$ in \eqref{eq:average_actuation_times} is replaced by $\|u_k\|_{\ell_2/\ell_1}$, known as the $\ell_2/\ell_1$ norm \cite{eldar2010block}.
To ensure robustness against disturbances, we employ an MPC scheme with a prediction horizon of 30 and solve the $\ell_1$-relaxed problem at each time step using CVX \cite{grant2014cvx}.
This method is hereafter referred to as $\ell_1$-relax.+MPC.
We conduct Monte Carlo simulations for each method using different values of the weighting parameter $\theta \in \{0.02, 0.04, \dots, 0.40\}$, and compare the means of the control cost and the actuation rate across trials for each $\theta$.

\begin{figure}[t]
  \centering
  \includegraphics[width=\linewidth]{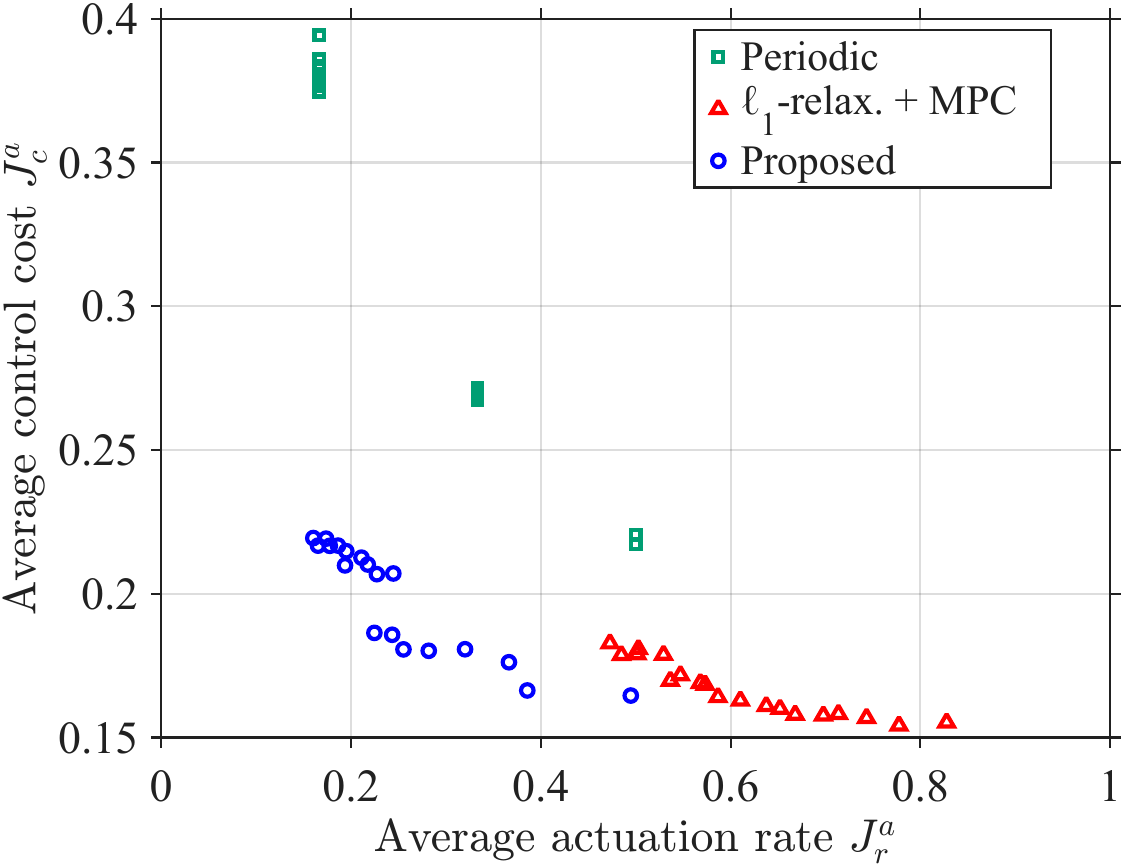}
  \caption{Plot of the average control cost versus the average actuation rate.}
  \label{fig:Jr_vs_Jc}
\end{figure}

\begin{figure}[t]
  \centering
  \includegraphics[width=\linewidth]{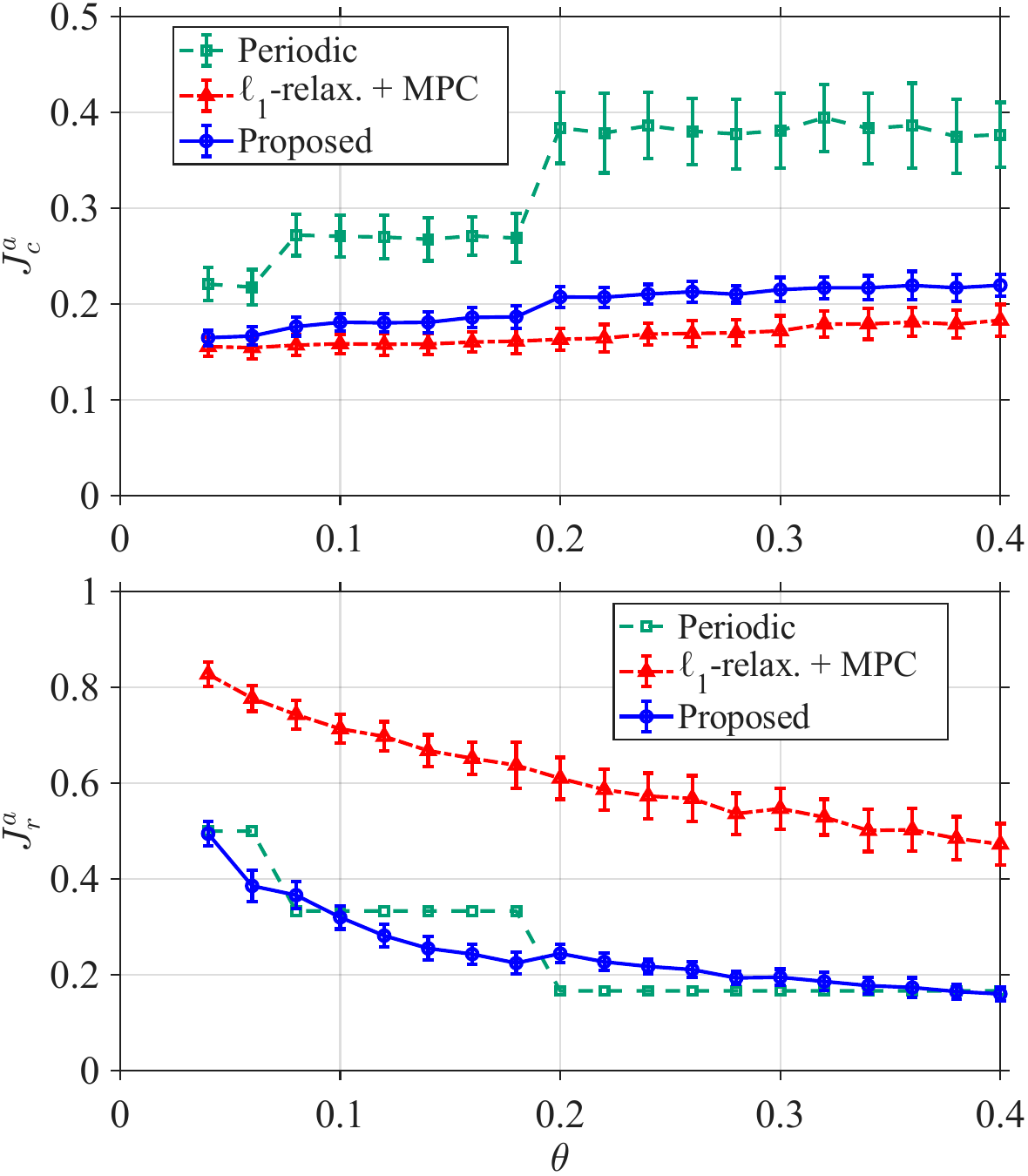}
  \caption{Average control cost (top) and average actuation rate (bottom) as functions of $\theta$, with error bars indicating the standard deviation from the mean.}
  \label{fig:theta_curves}
\end{figure}

Fig.~\ref{fig:Jr_vs_Jc} shows the average control cost versus the average actuation rate, and Fig.~\ref{fig:theta_curves} presents each cost as a function of $\theta$.
In Fig.~\ref{fig:theta_curves}, error bars indicate one standard deviation around the mean over 50 trials.
For the periodic control, $p \in \{1,2,3,6\}$ is chosen so that the average cost associated with the periodic policy \eqref{eq:optimal_periodic_cost_average} is minimized for each $\theta$.
Compared with periodic control, the proposed algorithm achieves lower control costs for all $\theta$, while maintaining nearly the same average actuation rate.
Although $\ell_1$-relax.+MPC yields better control performance than the proposed method, its average actuation rate is relatively high.
Thus, our method provides a superior trade-off compared with the other two approaches, highlighting the effectiveness of the rollout-based algorithm we propose.

\section{Conclusion}
\label{sec:conclusion}

We have proposed a sparse control framework that aims to balance the trade-off between control performance and actuation rate.
We formulated the optimal control problem using actuation-rate regularization and developed a rollout-based algorithm to obtain a tractable solution.
We established theoretical guarantees, showing that the proposed method achieves a performance guarantee with respect to periodic control and ensures mean-square stability of the closed-loop system.
The effectiveness of the approach was demonstrated through a numerical example.
Future work includes investigating performance guarantees for discounted-cost formulations and deriving strict performance improvement results over periodic control strategies.

\appendix

This appendix provides some preliminaries as well as the detailed proofs of the analytical results established in this paper.



Before proceeding to proofs, we derive the dynamics of $\hat{\chi}_\ell$ induced by the policies $(\mu^{u, \text{ro}}, \mu^{\delta, \text{ro}})$.
From \eqref{eq:estimation}, we have
\begin{align}
  &\hat{x}_{k+1} = (A + \rho_{k-\ell h}^{(\varphi(\hat{\chi}_\ell))} B F_{k-\ell h}^{(\varphi(\hat{\chi}_\ell))}) \hat{x}_k \\
  &\quad + G\left(y_{k+1} - C(A + \rho_{k-\ell h}^{(\varphi(\hat{\chi}_\ell))} B F_{k-\ell h}^{(\varphi(\hat{\chi}_\ell))}) \hat{x}_k\right). \label{eq:estimation_rewritten}
\end{align}
For $i \in \{0,1,\dots,h-1\}$ and $m \in \Mset$, we define
\begin{align}
  \Theta_i^{(m)} &\coloneqq A + \rho_i^{(m)} B F_i^{(m)}, \\
  \omega_k &\coloneqq G\left(CAe_k + Cw_k + v_{k+1}\right).
\end{align}
Then, Equation~\eqref{eq:estimation_rewritten} can be rewritten as
\begin{equation}
  \hat{x}_{k+1} = \Theta_{k-lh}^{(\varphi(\hat{\chi}_\ell))} \hat{x}_k + \omega_k.
\end{equation}
By iterating this equation from $k = \ell h$ to $k = (\ell+1)h-1$, we obtain
\begin{equation}
  \hat{\chi}_{\ell+1} = \Phi_{\varphi(\hat{\chi}_\ell)} \hat{\chi}_\ell + \Gamma_{\varphi(\hat{\chi}_\ell)} \bar{\omega}_\ell, \quad \forall \ell \in \Z_{\geq 0},
\end{equation}
where
\begin{align}
  \Phi_m &\coloneqq \prod_{s=h-1}^{0} \Theta_s^{(m)}, \\
  \Gamma_m &\coloneqq \begin{bmatrix}
    \prod_{s=h-1}^{1} \Theta_s^{(m)} & \prod_{s=h-1}^{2} \Theta_s^{(m)} & \dots & \Theta_{h-1}^{(m)} & I_{n_x}
  \end{bmatrix}, \\
  \bar{\omega}_\ell &\coloneqq \begin{bmatrix}
    \omega_{\ell h}^{\T} & \omega_{\ell h+1}^{\T} & \dots & \omega_{(\ell+1)h-1}^{\T}
  \end{bmatrix}^{\T}.
\end{align}


\subsection{Preliminary Lemmas}
\label{append:preliminary_lemmas}

\begin{lemma}
  \label{lemma:observability_lifted_system}
  If $(A,Q^{1/2})$ is observable, then for any $p \in \N$, the pair $(A^p, Q_p^{1/2})$ is also observable.
\end{lemma}
\begin{proof}
  Let $\lambda_p \in \C$ be an eigenvalue of $A^p$ and $\xi_p \in \C^{n_x} \setminus \{0\}$ be a corresponding eigenvector, i.e., $A^p \xi_p = \lambda_p \xi_p$.
  Suppose that there exists such a vector $\xi_p$ satisfying $Q_p^{1/2} \xi_p = 0$.
  From the definition of $Q_p$, it follows that
  \begin{equation}
    0 = \xi_p^\ast Q_p \xi_p = \sum_{i=0}^{p-1} (A^i \xi_p)^\ast Q A^i \xi_p = \sum_{i=0}^{p-1} \| Q^{1/2} A^i \xi_p \|^2.
  \end{equation}
  Hence, $Q^{1/2} A^i \xi_p = 0$ for all $i = 0, 1, \dots, p-1$.
  Writing $s = qp + r$ with $q \in \N$ and $r \in \{0, 1, \dots, p-1\}$, we obtain
  \begin{equation}
    \label{eq:eigenvalue_decomposition}
    A^s \xi_p = A^{qp + r} \xi_p = \lambda_p^q A^r \xi_p.
  \end{equation}
  Since $Q^{1/2} A^r \xi_p = 0$ for all $r = 0, 1, \dots, p-1$, Equation~\eqref{eq:eigenvalue_decomposition} implies that $Q^{1/2} A^s \xi_p = 0$ for all $s \in \N$.
  Consequently,
  \begin{equation}
    \xi_p \in \{x \in \C^{n_x} \colon Q^{1/2} A^s x = 0, \; \forall s \in \N\}.
  \end{equation}
  On the other hand, since $(A,Q^{1/2})$ is observable, the Cayley--Hamilton theorem ensures that
  \begin{equation}
    \{x \in \C^{n_x} \colon Q^{1/2} A^s x = 0, \; \forall s \in \N\} = \{0\}.
  \end{equation}
  This contradicts $\xi_p \neq 0$.
  Therefore, $Q_p^{1/2} \xi_p \neq 0$ holds for all eigenvectors $\xi_p$ of $A^p$, which completes the proof.
\end{proof}

\begin{lemma}
  \label{lemma:positive_definiteness_Pp}
  Suppose that $(A,Q^{1/2})$ is observable and that Assumption~\ref{assump:eigenvalue_condition} holds for some $p \in \N$.
  Then, the matrix $\tilde{P}_p$ obtained by solving \eqref{eq:algebraic_riccati_discounted} with $\alpha=1$ is positive definite.
\end{lemma}
\begin{proof}
  Assumption~\ref{assump:eigenvalue_condition} and Lemma~\ref{lemma:observability_lifted_system} ensure that $\tilde{P}_p$ is positive semidefinite at $\alpha = 1$ \cite[Sec.~8.4]{bertsekas1976dynamic}.
  Assume that there exists a nonzero vector $\xi \in \R^{n_x}$ such that $\xi^\top \tilde{P}_p \xi = 0$.
  Then, we consider the following optimal control problem for a noise-free system with the initial condition $x_0 = \xi$:
  \begin{mini}|s|
    {\{u_k\}_{k \in \Z_{\geq 0}}}
    {\sum_{k=0}^{\infty} x_k^\T Q_p x_k + 2x_k^{\T} S_p u_k + u_k^{\T} R_p u_k}
    {\label{prob-append:optimal_control_problem}}{}
    {}
    \addConstraint{x_{k+1}}{= A^p x_k + B_p u_k, \quad k \in \Z_{\geq 0}}
    \addConstraint{x_0}{= \xi.}
  \end{mini}
  Since $(A^p, B_p)$ is stabilizable and $(A^p, Q_p^{1/2})$ is observable by Lemma~\ref{lemma:observability_lifted_system}, the optimal value of \eqref{prob-append:optimal_control_problem} is $\xi^\top \tilde{P}_p \xi = 0$.
  Furthermore, by the definitions of $Q_p$, $S_p$, and $R_p$, the stage cost in \eqref{prob-append:optimal_control_problem} can be rewritten as
  \begin{align}
    &x_k^\T Q_p x_k + 2x_k^{\T} S_p u_k + u_k^{\T} R_p u_k \\
    &= u_k^\T R u_k + x_k^\T Q x_k \\
    &\quad + \sum_{i=1}^{p-1} \left(A^i x_k + A^{i-1} B u_k\right)^\T Q \left(A^i x_k + A^{i-1} B u_k\right) \\
    &\geq 0,
  \end{align}
  where the inequality follows from $Q \succeq 0$ and $R \succ 0$.
  Hence, the equality
  \begin{equation}
    x_k^\T Q_p x_k + 2x_k^{\T} S_p u_k + u_k^{\T} R_p u_k = 0,\quad \forall k \in \Z_{\geq 0}
  \end{equation}
  implies $u_k^\T R u_k = 0$, and thus $u_k = 0$ for all $k \in \Z_{\geq 0}$.
  Substituting $u_k=0$ into the above equality yields $x_k^\T Q_p x_k = 0$, and hence $ Q_p^{1/2} x_k = 0$ for all $k \in \Z_{\geq 0}$.
  Since $u_k = 0$ for all $k \in \Z_{\geq 0}$, it holds that $x_k = A^{pk} \xi$ for all $k \in \Z_{\geq 0}$.
  Consequently, $Q_p^{1/2} A^{pk} \xi = 0$ for all $k \in \Z_{\geq 0}$, which contradicts the observability of $(A^p, Q_p^{1/2})$ established in Lemma~\ref{lemma:observability_lifted_system}.
  Therefore, only $\xi = 0$ satisfies $\xi^\top \tilde{P}_p \xi = 0$, and thus $\tilde{P}_p$ is positive definite at $\alpha = 1$.
\end{proof}

\begin{lemma}
  \label{lemma:initial_value_P0}
  Consider Algorithm~1 under Assumptions~\ref{assump:eigenvalue_condition} and \ref{assump:performance_and_stability_guarantees}.  
  Then, it holds that $P^{(1)}_0 = \tilde{P}_p$.
\end{lemma}
\begin{proof}
  First, we can observe that $\hat{V}^d_k$ in \eqref{eq:rollout_cost} is obtained by applying the control policy $\mu^u_{\text{ro}}$ to the $h$-stage cost \eqref{eq:cost_to_minimize}, starting from $k = \ell h$ for $\ell \in \Z_{\geq 0}$.

  Define the following function:
  \begin{align}
    J_h(\Iset_k) &\coloneqq \E\Biggl[x_{k+h}^\T \tilde{P}_p x_{k+h} \\
    &\quad + \sum_{i=k}^{k+h-1} \bigl(x_i^\T Q x_i + u_i^\T R u_i + \theta \delta_i \bigr) \,\bigg|\, \Iset_k\Biggr].
  \end{align}
  Then, we consider the following $h$-stage cost minimization problem:
  \begin{mini}|s|
    {\{u_i\}_{i=k}^{k+h-1}, \{\delta_i\}_{i=k}^{k+h-1}}
    {J_h(\Iset_k)}
    {\label{prob-append:h_stage_cost_minimization}}{}
    {}
    \addConstraint{x_{i+1}}{= A x_i + B u_i + w_i}
    \addConstraint{y_i}{= C x_i + v_i}
    \addConstraint{\delta_i}{\in \{0,1\}}
    \addConstraint{u_i}{= 0 \quad \text{if } \delta_i = 0.}
  \end{mini}
  Let $H \coloneqq h/p \in \N$.  
  If the periodic triggering policy with period $p$ is optimal for \eqref{prob-append:h_stage_cost_minimization}, then $J_h$ can be rewritten as
  \begin{align}
    J_h(\Iset_k) &= \E\Biggl[x_{k+h}^\T \tilde{P}_p x_{k+h} + \sum_{j=0}^{H-1} \sum_{i=0}^{p-1} x_{k+jp+i}^\T Q x_{k+jp+i} \nonumber \\
    &\qquad + u_{k+jp+i}^\T R u_{k+jp+i} + \theta \delta_{k+jp+i} \,\Big|\, \Iset_k\Biggr] \\
    &= \E\Biggl[x_{k+h}^\T \tilde{P}_p x_{k+h} + \sum_{j=0}^{H-1}  x_{k+jp}^\T Q_p x_{k+jp} \nonumber \\
    &\qquad + 2x_{k+jp}^\T S_p u_{k+jp} + u_{k+jp}^\T R_p u_{k+jp} \,\Big|\, \Iset_k\Biggr] \\
    &\quad + H d^a(p) + H\theta.
  \end{align}
  By applying the dynamic programming principle, we obtain the following recursion:
  \begin{align}
    P_h &= \tilde{P}_p, \\
    P_t &= Q_p + (A^p)^\T P_{t+1} A^p - \left((A^p)^\T P_{t+1} B_p + S_p \right) \\
    &\qquad \times \left(B_p^\T P_{t+1} B_p + R_p\right)^{-1} \left(B_p^\T P_{t+1} A^p + S_p^\T \right), \nonumber \\
    &\quad t \in \{0,1,\dots,H-1\}. \label{eq:riccati_equation_m_equal_1}
  \end{align}
  Comparing \eqref{eq:riccati_equation_m_equal_1} with \eqref{eq:algebraic_riccati_discounted} under $\alpha=1$ shows that $P_t = \tilde{P}_p$ for all $t \in \{0,1,\dots,H-1\}$.
  Therefore, the optimal value of \eqref{prob-append:h_stage_cost_minimization} corresponding to the periodic policy is given by
  \begin{equation}
    \label{eq:cost_with_m_equal_1}
    J_h(\Iset_k) = \hat{x}_k^\T \tilde{P}_p \hat{x}_k + \tr\left(\tilde{P}_p \Sigma\right) + \beta^{(1)} + \gamma^{(1)},
  \end{equation}
  where
  \begin{align}
    \beta^{(1)} &= H\Bigl( \tr\left(\tilde{P}_p D_p \Omega_w^{(p)} D_p^\T\right) + \tr\left(M_p \Sigma\right) + d^a(p) \Bigr), \label{eq-append:beta1} \\
    \gamma^{(1)} &= H\theta. \label{eq-append:gamma1}
  \end{align}
  Comparing \eqref{eq:rollout_cost} and \eqref{eq:cost_with_m_equal_1}, we obtain $P^{(1)}_0 = \tilde{P}_p$.
\end{proof}

\begin{lemma}
  \label{lemma:coerciveness}
  The function $V_1$ defined in \eqref{eq:foster_lyapunov_function} is coercive; that is, $V_1(x) \to \infty$ as $\|x\| \to \infty$.
\end{lemma}
\begin{proof}
  To establish the coerciveness of $V_1$, it is sufficient to show that the sublevel sets $\Lset_r(x) \coloneqq \{x \in \R^{n_x} \colon V_1(x) \leq r\}$ are precompact, i.e., their closures are compact, for each $r \in \R_{>0}$ \cite[Sec.~9.4]{meyn2012markov}.
  Since $P^{(1)}_0$ is positive definite by Lemmas~\ref{lemma:positive_definiteness_Pp} and \ref{lemma:initial_value_P0}, we have
  \begin{equation}
    \Lset_r(x) \subseteq \left\{ x \in \R^{n_x} \colon \|x\|^2 \leq \frac{r}{\lambda_{\min}(P^{(1)}_0)} \right\}.
  \end{equation}
  The set on the right-hand side is bounded for every $r \in \R_{>0}$, and hence $\Lset_r(x)$ is also bounded.
  Moreover, the continuity of $V_1$ implies that $\Lset_r(x) = V_1^{-1}([0,r])$ is closed \cite[Th.~4.8]{rudin1976principles}.
  By the Heine--Borel theorem \cite[Th.~2.41]{rudin1976principles}, $\Lset_r(x)$ is compact, and thus its closure is also compact for each $r \in \R_{>0}$ \cite[Th.~2.27]{rudin1976principles}, which establishes the claim.
\end{proof}

\subsection{Proof of Lemma~\ref{lemma:markov_chain}}
\label{append:markov_chain}

Let $\Borel(\R^{n_x})$ denote the Borel field generated by all open sets in $\R^{n_x}$, and define $\Fil_\ell \coloneqq \sigma(\hat{\chi}_0,\hat{\chi}_1,\dots,\hat{\chi}_\ell)$.
In the following, we first construct a function $P(x,\Aset)$ for $x \in \R^{n_x}$ and $\Aset \in \Borel(\R^{n_x})$, and show that $P$ is a transition probability kernel.
Then, we show that the process $\{\hat\chi_\ell\}_{\ell \in \Z_{\geq 0}}$ has the Markov property \cite[Th.~6.3.1]{durrett2010probability}.

We first show that the mapping $\varphi$ is measurable, which allow us to define a transition probability kernel.
For each $m \in \Mset$, we define $\Cset_m \coloneqq \{x \in \R^{n_x} \colon \varphi(x)=m\}$ and
\begin{equation}
  g_m(x)
  \coloneqq
  x^\top P^{(m)}_0 x + \tr(P^{(m)}_0\Sigma) + \beta^{(m)} + \gamma^{(m)}.
\end{equation}
Since $\varphi$ is determined by the minimization of finitely many continuous functions, we have
\begin{align}
  \Cset_m
  &=
  \left(\bigcap_{i < m} \{x \in \R^{n_x} \colon g_m(x) < g_i(x)\}\right) \\
  &\qquad \cap
  \left(\bigcap_{i > m} \{x \in \R^{n_x} \colon g_m(x) \le g_i(x)\}\right).
\end{align}
Since each $g_i-g_m$ is continuous, the sets $\{x \in \R^{n_x} \colon g_m(x) < g_i(x)\}$ and $\{x \in \R^{n_x} \colon g_m(x) \le g_i(x)\}$ are Borel measurable
\cite[Th.~4.8]{rudin1976principles}.
Hence, $\Cset_m \in \Borel(\R^{n_x})$ for all $m \in \Mset$.
Since $\varphi^{-1}(\{m\})=\Cset_m$ for each $m \in \Mset$, the mapping $\varphi$ is measurable.

Next, we construct the candidate transition probability kernel.
We define
\begin{align}
  \Omega_\omega
  &\coloneqq
  G \bigl(CA\Sigma A^\top C^\top + C\Omega_w C^\top + \Omega_v\bigr) G^\top, \\
  \bar\Omega_\omega
  &\coloneqq
  I_h \otimes \Omega_\omega.
\end{align}
Since $\Omega_v \succ 0$ and
Assumption~\ref{assump:performance_and_stability_guarantees}(i) holds, we have $\Omega_\omega \succ 0$ and $\bar\Omega_\omega \succ 0$.
Moreover, $\Gamma_m$ has full row rank for every $m \in \Mset$ by construction.
Therefore,
\begin{equation}
  \Sigma_m = \Gamma_m \bar\Omega_\omega \Gamma_m^\T \succ 0, \quad \forall m \in \Mset.
\end{equation}
For each $m \in \Mset$, define
\begin{equation} \label{eq-append:gaussian_distribution}
  p_m(x,z)
  \coloneqq
  \frac{
    \exp\left(
      -\frac12 (z-\Phi_m x)^\top \Sigma_m^{-1}(z-\Phi_m x)
    \right)
  }{
    (2\pi)^{n_x/2}\sqrt{\det(\Sigma_m)}
  }.
\end{equation}
Fix any $x \in \R^{n_x}$ and let $m=\varphi(x)$.
Since $\bar{\omega}_\ell$ is Gaussian with mean zero and covariance matrix $\bar\Omega_\omega$,
the random vector $\Phi_m x + \Gamma_m \bar{\omega}_\ell$ is Gaussian with mean $\Phi_m x$ and covariance matrix $\Sigma_m$.
Hence, the function $p_{\varphi(x)}(x,\cdot)$ given by \eqref{eq-append:gaussian_distribution}
is the natural candidate for the density of the transition function from the state $x$, and we define the function $P$ on $\R^{n_x}\times\Borel(\R^{n_x})$ by
\begin{equation} \label{eq-append:transition_kernel}
  P(x,\Aset)
  \coloneqq
  \int_\Aset p_{\varphi(x)}(x,z)\,dz,
  \quad
  x \in \R^{n_x},
  \ \Aset \in \Borel(\R^{n_x}).
\end{equation}

We show that $P$ given by \eqref{eq-append:transition_kernel} is a transition probability kernel by proving that $\Aset \mapsto P(x,\Aset)$ is a probability measure for each $x \in \R^{n_x}$ and $x \mapsto P(x,\Aset)$ is a nonnegative measurable function for each $\Aset \in \Borel(\R^{n_x})$ \cite[Sec.~3.4]{meyn2012markov}.
For each fixed $x \in \R^{n_x}$, the mapping
\begin{equation}
  \Aset \mapsto P(x,\Aset)=\int_\Aset p_{\varphi(x)}(x,z)\,dz
\end{equation}
is a probability measure on $\Borel(\R^{n_x})$ since $p_{\varphi(x)}(x,\cdot)$ is the density of a Gaussian distribution, and a distribution induces a probability measure on Borel sets \cite[Sec.~1.2]{durrett2010probability}.
Next, fix any $\Aset \in \Borel(\R^{n_x})$.
Since each $p_m$ is continuous on $\R^{n_x}\times\R^{n_x}$, it is Borel measurable \cite[Th.~4.8]{rudin1976principles}.
Moreover,
\begin{equation}
  p_{\varphi(x)}(x,z)
  =
  \sum_{m \in \Mset}
  \mathbb{I}_{\Cset_m}(x)\,p_m(x,z)
\end{equation}
is Borel measurable on $\R^{n_x}\times\R^{n_x}$ since $\Mset$ is finite, $\Cset_m \in \Borel(\R^{n_x})$ for all $m \in \Mset$, $\mathbb{I}_{\Cset_m}$ is measurable \cite[Def.~11.19]{rudin1976principles}, and finite sums and products of measurable functions are also measurable \cite[Th.~11.18]{rudin1976principles}.
Therefore, the mapping
\begin{equation}
  x
  \mapsto
  P(x,\Aset)
  =
  \int_{\R^{n_x}}
  \mathbb{I}_\Aset(z)\,p_{\varphi(x)}(x,z)\,dz
\end{equation}
is Borel measurable by Fubini's theorem \cite[Th.~1.7.2]{durrett2010probability}.
Hence, $P$ given by \eqref{eq-append:transition_kernel} is a transition probability kernel.

We finally prove the Markov property of $\{\hat\chi_\ell\}_{\ell \in \Z_{\geq 0}}$.
By construction, $\bar{\omega}_\ell$ is a measurable function of $e_{\ell h}$ and $\{w_k,v_{k+1} \colon \ell h \le k \le (\ell+1)h-1\}$.
Moreover, since $e_{\ell h}$ is Gaussian and orthogonal to $L^2(\sigma(\Iset_{\ell h}))$, it is independent of $\sigma(\Iset_{\ell h})$, and the noises $\{w_k,v_{k+1} \colon \ell h \le k \le (\ell+1)h-1\}$ are also independent of $\sigma(\Iset_{\ell h})$.
Therefore, it holds that $\sigma(\bar{\omega}_\ell)\perp \sigma(\Iset_{\ell h})$.
Since $\Fil_\ell \subseteq \sigma(\Iset_{\ell h})$, it follows that $\sigma(\bar{\omega}_\ell)\perp \Fil_\ell$ and $\sigma(\bar{\omega}_\ell)\perp \sigma(\hat{\chi}_\ell)$.
Fix any $\Aset \in \Borel(\R^{n_x})$ and define $\Xi_\ell \coloneqq (\hat{\chi}_0,\hat{\chi}_1,\dots,\hat{\chi}_\ell)$.
Then $\Fil_\ell=\sigma(\Xi_\ell)$.
Since each mapping $(x,\omega)\mapsto \Phi_mx+\Gamma_m\omega$ is continuous and $\varphi$ is measurable, the mapping $(\xi,\omega) \mapsto \mathbb{I}_\Aset\left( \Phi_{\varphi(\xi_\ell)}\xi_\ell+\Gamma_{\varphi(\xi_\ell)}\omega\right)$ is Borel measurable, since $\mathbb{I}_{\Cset_m}$ is Borel measurable and $\Phi_{\varphi(\xi_\ell)}\xi_\ell+\Gamma_{\varphi(\xi_\ell)}\omega$ is measurable \cite[Th.~1.12]{rudin1987real}.
Since $\sigma(\bar{\omega}_\ell)\perp \Fil_\ell$ means $\Xi_\ell$ and $\bar{\omega}_\ell$ are independent, \cite[Ex.~4.1.7]{durrett2010probability} gives
\begin{align}
  &\Prob\left(\hat{\chi}_{\ell+1}\in\Aset \midd \Fil_\ell\right) \\
  &=
  \E\left[
    \mathbb{I}_\Aset\left(
      \Phi_{\varphi(\hat{\chi}_\ell)}\hat{\chi}_\ell
      +
      \Gamma_{\varphi(\hat{\chi}_\ell)}\bar{\omega}_\ell
    \right)
    \midd
    \Xi_\ell
  \right] \\
  &=
  \int_{\R^{n_x h}}
  \mathbb{I}_\Aset\left(
    \Phi_{\varphi(\hat{\chi}_\ell)}\hat{\chi}_\ell
    +
    \Gamma_{\varphi(\hat{\chi}_\ell)}\omega
  \right)
  \nu_{\bar{\omega}}(d\omega) \\
  &=
  P(\hat{\chi}_\ell,\Aset)
  \quad \text{a.s.} \label{eq-append:P1}
\end{align}
From $\sigma(\bar{\omega}_\ell)\perp \sigma(\hat{\chi}_\ell)$, applying \cite[Ex.~4.1.7]{durrett2010probability} again yields
\begin{equation}
  \Prob\left(\hat{\chi}_{\ell+1}\in\Aset \midd \hat{\chi}_\ell\right)
  =
  P(\hat{\chi}_\ell,\Aset)
  \quad \text{a.s.} \label{eq-append:P2}
\end{equation}
By combining \eqref{eq-append:P1} and \eqref{eq-append:P2}, we obtain
\begin{equation}
  \Prob\left(\hat{\chi}_{\ell+1}\in\Aset \midd \Fil_\ell\right)
  =
  \Prob\left(\hat{\chi}_{\ell+1}\in\Aset \midd \hat{\chi}_\ell\right)
  \quad \text{a.s.}
\end{equation}
Therefore, the process $\{\hat{\chi}_\ell\}_{\ell \in \Z_{\geq 0}}$ is a time-homogeneous
Markov chain with transition probability kernel $P$ \cite[Sec.~3.4]{meyn2012markov}.
\hfill\QED

\subsection{Proof of Lemma~\ref{lemma:foster_lyapunov}}
\label{append:foster_lyapunov}

We define a function
\begin{align}
  &\bar{g}(\chi_\ell, \hat{\chi}_\ell, \bar{w}_\ell, \bar{\omega}_\ell) \\
  &\coloneqq \sum_{\tau=0}^{h-1} g\left(x_{\ell h+\tau},\; \rho^{(\varphi(\hat{\chi}_\ell))}_\tau F^{(\varphi(\hat{\chi}_\ell))}_\tau \hat{x}_{\ell h+\tau},\; \rho^{(\varphi(\hat{\chi}_\ell))}_\tau\right) \label{eq:def_barf}
\end{align}
and $\bar{w}_\ell \coloneqq [w_{\ell h}^\T\ w_{\ell h+1}^\T\ \dots w_{(\ell+1)h-1}^\T]^\T$.

First, we show that there exists a sufficiently small constant $a_1>0$ such that
\begin{equation}
  \label{eq-append:inequality_a1}
  \E\left[\bar{g}(\chi_\ell, \hat{\chi}_\ell, \bar{w}_\ell, \bar{\omega}_\ell) \midd \hat{\chi}_\ell\right] \geq a_1 \hat{\chi}_\ell^\T \hat{\chi}_\ell,\quad \forall \ell \in \Z_{\geq 0}.
\end{equation}
Since $R \succ 0$ and $\theta > 0$, it follows that
\begin{align}
  &\E\left[\bar{g}(\chi_\ell, \hat{\chi}_\ell, \bar{w}_\ell, \bar{\omega}_\ell) \midd \hat{\chi}_\ell\right] \\
  &\geq \E\left[ \sum_{\tau=0}^{h-1} x_{\ell h+\tau}^\T Q x_{\ell h+\tau} \midd \hat{\chi}_\ell\right]
  \geq \E\left[x_{\ell h}^\T Q x_{\ell h} \midd \hat{\chi}_\ell\right] \\
  &= \E\left[(\hat\chi_\ell + e_{\ell h})^\T Q (\hat\chi_\ell + e_{\ell h}) \midd \hat{\chi}_\ell\right]
  =\hat\chi_\ell^\T Q \hat\chi_\ell + \tr(Q\Sigma) \\
  &\geq \lambda_{\min}(Q) \hat\chi_\ell^\T \hat\chi_\ell.
\end{align}
From Assumption~\ref{assump:performance_and_stability_guarantees}(ii), we can set $a_1 = \lambda_{\min}(Q) > 0$, which satisfies \eqref{eq-append:inequality_a1}.

Furthermore, for any constant $b_1$ such that $b_1 > a_1 > 0$ and $b_1I_{n_x} - \tilde{P}_p \succ 0$, we obtain
\begin{equation}
  \E\left[V_1(\chi_\ell) \midd \hat{\chi}_\ell\right]
  \leq
  b_1\hat{\chi}_\ell^\T \hat{\chi}_\ell + b_1\tr\left(\Sigma\right)
\end{equation}
from \eqref{eq:foster_lyapunov_function}.

Hence, we have
\begin{align}
  \E\left[V_1(\chi_{\ell+1}) \midd \hat{\chi}_\ell\right]
  &\leq \beta^{(1)} + \gamma^{(1)} + \E\left[V_1(\chi_\ell) \midd \hat{\chi}_\ell\right] \\
  &\quad - a_1\hat{\chi}_\ell^\T \hat{\chi}_\ell \\
  &\leq d_1 \E\left[V_1(\chi_\ell) \midd \hat{\chi}_\ell\right] + c_1, \label{eq:recursion_V1}
\end{align}
where
\begin{align}
  d_1 &= 1 - \frac{a_1}{b_1} < 1, \\
  c_1 &= \beta^{(1)} + \gamma^{(1)} + a_1 \tr\left(\Sigma\right).
\end{align}
Let $\kappa \coloneqq \tr(P^{(1)}_0\Sigma)$.
Then
\begin{align}
  &\E\left[V_1(\chi_\ell) \midd \hat{\chi}_\ell\right] = V_1(\hat{\chi}_\ell) + \kappa, \\
  &\E\left[V_1(\chi_{\ell+1}) \midd \hat{\chi}_\ell\right] = \E\left[V_1(\hat{\chi}_{\ell+1}) \midd \hat{\chi}_\ell\right] + \kappa.
\end{align}
Substituting these equalities into \eqref{eq:recursion_V1} yields
\begin{equation}
  \E\left[V_1(\hat{\chi}_{\ell+1}) \midd \hat{\chi}_\ell\right]
  \le d_1 V_1(\hat{\chi}_\ell) + c_2,
\end{equation}
where $c_2 \coloneqq c_1 + (d_1-1)\kappa$.
Taking conditional expectation with respect to $\hat{\chi}_0$ and using the tower property and the Markov property of $\{\hat{\chi}_\ell\}_{\ell\in\Z_{\ge 0}}$, we obtain
\begin{equation}
  \E\left[V_1(\hat{\chi}_{\ell+1}) \midd \hat{\chi}_0\right]
  \le d_1 \E\left[V_1(\hat{\chi}_\ell) \midd \hat{\chi}_0\right] + c_2.
\end{equation}
Iterating this inequality for $\ell=0,1,\dots,L-1$ gives
\begin{equation}
  \E\left[V_1(\hat{\chi}_L) \midd \hat{\chi}_0\right]
  \le d_1^L V_1(\hat{\chi}_0) + \sum_{i=0}^{L-1} d_1^i c_2.
\end{equation}
Finally, by the tower property, we have
\begin{align}
  \E\left[V_1(\chi_L) \midd \hat{\chi}_0\right]
  &=
  \E\left[\E\left[V_1(\chi_L) \midd \hat{\chi}_L\right] \midd \hat{\chi}_0\right] \\
  &=
  \E\left[V_1(\hat{\chi}_L) \midd \hat{\chi}_0\right] + \kappa,
\end{align}
which shows that $\E\left[V_1(\chi_L) \midd \hat{\chi}_0\right]$ remains bounded as $L \to \infty$ since $d_1 < 1$.
\hfill\QED

\subsection{Proof of Theorem~\ref{thm:ergodicity}}
\label{append:proof_of_ergodicity}

In order to prove Theorem~\ref{thm:ergodicity}, we establish the following three lemmas:

\begin{lemma}
  \label{lemma:open_set_irreducibility}
  The Markov chain $\{\hat{\chi}_\ell\}_{\ell \in \Z_{\geq 0}}$ is open set irreducible and aperiodic.
\end{lemma}

\begin{lemma}
  \label{lemma:T-chain}
  Let $\Bset_\varepsilon(z) \coloneqq \{x \in \R^{n_x} \colon \|z-x\| < \varepsilon\}$ denote the $\varepsilon$-neighborhood of $z$ for a constant $\varepsilon > 0$.
  Then, for any $z \in \R^{n_x}$ and $\Aset \in \Borel(\R^{n_x})$, there exists a nonnegative continuous function $z \mapsto T(z, \Aset)$ such that, for any $x \in \Bset_\varepsilon(z)$,
  \begin{equation}
    P(x,\Aset) \geq T(x,\Aset),\quad 0 < T(x,\R^{n_x}) \leq 1.
  \end{equation}
\end{lemma}

\begin{lemma}
  \label{lemma:drift_condition}
  The function $V_1$ satisfies the following drift condition:
  \begin{equation}
    \E\left[V_1(\hat{\chi}_{\ell+1}) \midd \hat{\chi}_\ell\right] - V_1(\hat{\chi}_\ell) \leq -1,\quad \forall \hat{\chi}_\ell \in \R^{n_x} \setminus \Dset,
  \end{equation}
  where
  \begin{equation}
    \label{eq-append:drift_set}
    \Dset \coloneqq \left\{\chi \in \R^{n_x} \colon a_1 \chi^\T \chi \leq \beta^{(1)} + \gamma^{(1)} + 1\right\}.
  \end{equation}
\end{lemma}

First, we prove Lemma~\ref{lemma:open_set_irreducibility}.
We recall that 
\begin{equation}
  P(x,\Aset)=\int_\Aset p_{\varphi(x)}(x,z)\,dz,
\end{equation}
and $p_m,\, m \in \Mset$ is defined in \eqref{eq-append:gaussian_distribution}.
Since $\Sigma_m \succ 0$, the mapping $z \mapsto p_m(x,z)$ is continuous and strictly positive on $\R^{n_x}$ \cite[Sec.~4.4.3]{meyn2012markov}.
Then, for any nonempty open set $\Oset \subset \R^{n_x}$, we have
\begin{equation}
  P(x,\Oset)=\int_{\Oset} p_{\varphi(x)}(x,z)\,dz > 0,
  \quad \forall x \in \R^{n_x},
\end{equation}
since $p_{\varphi(x)}(x,z)>0$ for all $z\in\R^{n_x}$.
Therefore, the chain $\{\hat{\chi}_\ell\}_{\ell \in \Z_{\geq 0}}$ is open set irreducible \cite[Sec.~6.1.2]{meyn2012markov}.

Next, we prove aperiodicity.
Fix any closed ball $\mathcal{K} \subset \R^{n_x}$ with nonempty interior.
We note that this $\mathcal{K}$ is compact because of \cite[Th.~2.41]{rudin1976principles}.
It suffices to show that there exists a nontrivial measure $\nu$ on $\Borel(\R^{n_x})$ such that $P(x,\mathcal{B}) \geq \nu(\mathcal{B})$ holds for all $x \in \mathcal{K}$ and $\mathcal{B} \in \Borel(\R^{n_x})$.
For each $m \in \Mset$, the mapping $(x,z)\mapsto p_m(x,z)$ is continuous and strictly positive on the compact set $\mathcal{K} \times \mathcal{K}$.
Hence, from \cite[Th.~4.46]{rudin1976principles}, there exists
\begin{equation} \label{eq-append:def_delta_m}
  \delta_m \coloneqq \min_{(x,z)\in \mathcal{K}\times \mathcal{K}} p_m(x,z) > 0.
\end{equation}
Since $\Mset$ is finite, the constant
\begin{equation}
  \delta \coloneqq \min_{m\in\Mset} \delta_m
\end{equation}
is well defined and $\delta>0$.
Define a measure $\nu$ on $\Borel(\R^{n_x})$ by
\begin{equation} \label{eq-append:def_nu}
  \nu(\mathcal{B})
  \coloneqq
  \delta \int_\mathcal{B} \mathbb{I}_\mathcal{K}(x)\, d\lambda(x)
  =
  \delta \lambda(\mathcal{B} \cap \mathcal{K}),
  \quad \mathcal{B}\in\Borel(\R^{n_x}),
\end{equation}
where $\lambda$ denotes the Lebesgue measure on $\Borel(\R^{n_x})$.
Then $\nu(\mathcal{K})=\delta\lambda(\mathcal{K})>0$.
Moreover, for any $x\in \mathcal{K}$ and $\mathcal{B}\in\Borel(\R^{n_x})$, we obtain
\begin{align}
  P(x,\mathcal{B})
  &= \int_{\mathcal{B}} p_{\varphi(x)}(x,z)\,dz
  \ge \int_{\mathcal{B}\cap \mathcal{K}} p_{\varphi(x)}(x,z)\,dz \\
  &\ge \delta \lambda(\mathcal{B} \cap \mathcal{K})
  = \nu(\mathcal{B}) \label{eq-appemd:minorization}
\end{align}
from \eqref{eq-append:def_delta_m} and \eqref{eq-append:def_nu}.
Equation \eqref{eq-appemd:minorization} implies that $\mathcal{K}$ is a small set \cite[Sec.~5.2]{meyn2012markov}.
Hence, the chain is aperiodic \cite[Sec.~5.4.3]{meyn2012markov}.

In the next step, we prove Lemma~\ref{lemma:T-chain}.
For any $\Aset \in \Borel(\R^{n_x})$, we define
\begin{align}
  &T(x,\Aset) \coloneqq \int_{\Aset} \underline p(x,z)\, dz,
  \quad \forall x \in \R^{n_x}, \label{eq:T_kernel} \\
  &\underline p(x,z) \coloneqq \min_{m \in \Mset} p_m(x,z)
\end{align}
Since $\Mset$ is finite, the minimum in \eqref{eq:T_kernel} always exists.
Moreover, for each fixed $x\in\R^{n_x}$, the map $z\mapsto p_m(x,z)$ is a measurable density,
and hence $z\mapsto \underline p(x,z)=\min_{m\in\Mset}p_m(x,z)$ is also measurable and nonnegative.
Therefore, $\Aset\mapsto T(x,\Aset)=\int_{\Aset}\underline p(x,z)\,dz$ is a measure on $\Borel(\R^{n_x})$ for all $x$ \cite[Sec.~1.4]{durrett2010probability}, which means that $\Aset\mapsto T(x,\Aset)$ is nonnegative.

We then show that $x \mapsto T(x,\Aset)$ is continuous.
Fix $\xi\in\R^{n_x}$.
For each $m\in\Mset$, we define $q_m(y) \coloneqq p_m(0,y)$ so that $p_m(x,z)=q_m(z-\Phi_m x)$ and $p_m(\xi,z)=q_m(z-\Phi_m \xi)$.
Then, by the translation continuity in $L^1$ \cite[Th.~9.5]{rudin1987real}, it holds that
\begin{align}
  &\|p_m(x,\cdot)-p_m(\xi,\cdot)\|_{L^1} \\
  &= \|q_m(\cdot-\Phi_m(x-\xi))-q_m(\cdot)\|_{L^1} \\
  &\to 0 \quad (x\to \xi). \label{eq:pm_convergence}
\end{align}
We prove the continuity of $\underline p(x,\cdot)$.
Using the inequality $|\min_i a_i-\min_i b_i| \leq \sum_i |a_i-b_i|$, we obtain, for all $z\in\R^{n_x}$,
\begin{align}
  &\left|\underline p(x,z)-\underline p(\xi,z)\right| \\
  &= \left|\min_{m\in\Mset} p_m(x,z)-\min_{m\in\Mset} p_m(\xi,z)\right| \\
  &\leq \sum_{m\in\Mset} |p_m(x,z)-p_m(\xi,z)|. \label{eq:pointwise_ineq}
\end{align}
Integrating both sides over $\R^{n_x}$ yields
\begin{align}
  &\int_{\R^{n_x}} \left|\underline p(x,z)-\underline p(\xi,z)\right|\, dz \\
  &\leq \sum_{m\in\Mset} \int_{\R^{n_x}} |p_m(x,z)-p_m(\xi,z)|\, dz. \label{eq:underline_p_int}
\end{align}
Since $p_m(x,z)$ is a nonnegative density for each $m \in \Mset$, we have
\begin{align}
  &\int_{\R^{n_x}} \left|p_m(x,z)-p_m(\xi,z)\right|\, dz \\
  &\leq \int_{\R^{n_x}} p_m(x,z)\, dz + \int_{\R^{n_x}} p_m(\xi,z)\, dz \\
  &= 1 + 1 < \infty,
\end{align}
which means that $p_m(x,z) - p_m(\xi,z) \in L^1$.
It also holds that $\underline p(x,z) - \underline p(\xi,z) \in L^1$ from \eqref{eq:pointwise_ineq}.
Hence, \eqref{eq:pm_convergence} and \eqref{eq:underline_p_int} result in
\begin{align}
  &\|\underline p(x,\cdot)-\underline p(\xi,\cdot)\|_{L^1} \\
  &\le \sum_{m\in\Mset} \|p_m(x,\cdot)-p_m(\xi,\cdot)\|_{L^1}
  \to 0 \quad (x\to \xi),
\end{align}
since $\Mset$ is finite.
Therefore, for any $\Aset\in\Borel(\R^{n_x})$,
\begin{align}
  |T(x,\Aset)-T(\xi,\Aset)|
  &= \left|\int_{\Aset} \left(\underline p(x,z)-\underline p(\xi,z)\right)\,dz\right| \\
  &\le \int_{\Aset} \left|\underline p(x,z)-\underline p(\xi,z)\right|\,dz \\
  &\le \|\underline p(x,\cdot)-\underline p(\xi,\cdot)\|_{L^1} \\
  &\to 0 \quad (x\to \xi),
\end{align}
which implies that $x\mapsto T(x,\Aset)$ is continuous.

For each $m \in \Mset$, we define
\begin{align}
  \tilde{P}_m(x,\Aset) &\coloneqq \Prob\left(\Phi_m x + \Gamma_m \bar{\omega}_0 \in \Aset\right) \\
  &= \nu_{\bar\omega} \left(\{\omega \colon \Phi_m x + \Gamma_m \omega \in \Aset\}\right). \label{eq:P_tilde}
\end{align}
By the definition of $P$, for any $\varepsilon>0$ and all $x\in \Bset_{\varepsilon}(z)$, we have
\begin{align}
  P(x,\Aset)
  &= \tilde{P}_{\varphi(x)}(x,\Aset)
  = \int_{\Aset} p_{\varphi(x)}(x,\zeta)\,d\zeta \\
  &\ge \int_{\Aset} \min_{m\in\Mset} p_m(x,\zeta)\,d\zeta
  = T(x,\Aset).
\end{align}
Moreover, since $p_m(x,z)>0$ for all $x,z$ and all $m\in\Mset$, we have $\underline p(x,z)>0$ for all $x,z$, and hence
\begin{equation}
  0 < T(x,\R^{n_x}) = \int_{\R^{n_x}} \underline p(x,z)\,dz \le 1,
\end{equation}
where the inequality follows from $\underline p(x,z)\le p_m(x,z)$ for any fixed $m\in\Mset$
and
\begin{equation}
  \int_{\R^{n_x}} \min_{m\in\Mset} p_m(x,z)\,dz \leq \int_{\R^{n_x}} p_m(x,z)\,dz = 1.
\end{equation}

Afterwards, we prove Lemma~\ref{lemma:drift_condition}.
Since $\hat{x}_k$ is a sufficient statistics with respect to $\Iset_k$, and by the orthogonal property of the Kalman filter, it follows that, for any $k \in \Z_{\geq 0}$,
\begin{align}
  \E\left[e_k^{\T} \hat{x}_k \midd \hat{x}_k\right] &= \E\left[(x_k - \hat{x}_k)^{\T} \hat{x}_k \midd \hat{x}_k\right] \\
  &= \left(\E\left[x_k\midd \hat{x}_k\right] -\hat{x}_k\right)^{\T} \hat{x}_k \\
  &= 0.
\end{align}
Hence, we obtain
\begin{align}
  \E\left[V_1(\chi_\ell) \midd \hat{\chi}_\ell\right] &= \E\left[(\hat{\chi}_\ell + e_{\ell h})^{\T} P^{(1)}_0 (\hat{\chi}_\ell + e_{\ell h}) \midd \hat{\chi}_\ell\right] \\
  &= \E\left[V_1(\hat{\chi}_\ell) \midd \hat{\chi}_\ell\right] + \tr\left(P^{(1)}_0 \Sigma\right),
\end{align}
and
\begin{align}
  &\E\left[V_1(\chi_{\ell+1}) \midd \hat{\chi}_\ell\right] \\
  &= \E\left[(\hat{\chi}_{\ell+1} + e_{(\ell+1)h})^{\T} P^{(1)}_0 (\hat{\chi}_{\ell+1} + e_{(\ell+1)h}) \midd \hat{\chi}_\ell\right] \\
  &= \E\left[V_1(\hat{\chi}_{\ell+1}) \midd \hat{\chi}_\ell\right] + \tr\left(P^{(1)}_0 \Sigma\right).
\end{align}
Therefore, we have
\begin{align}
  &\E\left[V_1(\chi_{\ell+1}) + \bar{g}\left(\chi_\ell, \hat{\chi}_\ell, \bar{w}_\ell, \bar{\omega}_\ell\right) \midd \hat{\chi}_\ell\right]  - \E\left[V_1(\chi_\ell) \midd \hat{\chi}_\ell\right] \\
  &= \E\left[V_1(\hat{\chi}_{\ell+1}) + \bar{g}\left(\chi_\ell, \hat{\chi}_\ell, \bar{w}_\ell, \bar{\omega}_\ell\right) \midd \hat{\chi}_\ell\right] - \E\left[V_1(\hat{\chi}_\ell) \midd \hat{\chi}_\ell\right] \\
  &= \beta^{(1)} + \gamma^{(1)} - f(\hat{\chi}_\ell),
\end{align}
where the first equality follows by using $\E[\chi_\ell^\top P^{(m)}_0\chi_\ell\mid \hat\chi_\ell]
= \hat\chi_\ell^\top P^{(m)}_0\hat\chi_\ell+\tr(P^{(m)}_0\Sigma)$ for all $m\in\Mset$, and the second equality follows from
\begin{align}
  &\E\left[V_1(\chi_{\ell+1}) + \bar{g}\left(\chi_\ell, \hat{\chi}_\ell, \bar{w}_\ell, \bar{\omega}_\ell\right) \midd \hat{\chi}_\ell\right] \\
  &= \hat{\chi}_\ell P^{(\varphi(\hat{\chi}_\ell))}_0 \hat{\chi}_\ell + \tr\left(P^{(\varphi(\hat{\chi}_\ell))}_0\Sigma\right) \\
  &\quad + \beta^{(\varphi(\hat{\chi}_\ell))} + \gamma^{(\varphi(\hat{\chi}_\ell))}, \\
  &\E\left[V_1(\hat{\chi}_\ell) \midd \hat{\chi}_\ell\right] = \hat{\chi}_\ell P^{(1)}_0 \hat{\chi}_\ell + \tr\left(P^{(1)}_0\Sigma\right).
\end{align}
Using \eqref{eq-append:inequality_a1}, we obtain
\begin{align}
  &\E\left[V_1(\hat{\chi}_{\ell+1}) \midd \hat{\chi}_\ell\right] - V_1(\hat{\chi}_\ell) \\
  &= \beta^{(1)} + \gamma^{(1)} - f(\hat{\chi}_\ell) - \E\left[\bar{g}\left(\chi_\ell, \hat{\chi}_\ell, \bar{w}_\ell, \bar{\omega}_\ell\right) \midd \hat{\chi}_\ell\right] \\
  &\leq \beta^{(1)} + \gamma^{(1)} - a_1 \hat{\chi}_\ell^{\T}\hat{\chi}_\ell.
\end{align}
From the definition of the set $\Dset$ in \eqref{eq-append:drift_set}, it follows that, for any $\hat{\chi}_\ell \in \R^{n_x}\setminus\Dset$,
\begin{equation}
  \E\left[V_1(\hat{\chi}_{\ell+1}) \midd \hat{\chi}_\ell\right] - V_1(\hat{\chi}_\ell) \leq -1.
\end{equation}

Building upon Lemmas~\ref{lemma:open_set_irreducibility}, \ref{lemma:T-chain}, and \ref{lemma:drift_condition}, we are ready to prove that the process $\{\hat{\chi}_\ell\}_{\ell \in \Z_{\geq 0}}$ is $\tilde{f}$-ergodic.
Lemma~\ref{lemma:T-chain} shows that the chain is a T-chain \cite[Prop.~6.2.4]{meyn2012markov}, and by combining this lemma and Lemma~\ref{lemma:open_set_irreducibility}, we can say $\{\hat{\chi}_\ell\}_{\ell \in \Z_{\geq 0}}$ is a $\psi$-irreducible T-chain \cite[Prop.~6.2.2]{meyn2012markov}.
Hence, every compact set in $\R^{n_x}$ is petite \cite[Th.~6.2.5]{meyn2012markov}, which yields that the set $\Dset$ is petite.
Then, Lemmas~\ref{lemma:coerciveness} and \ref{lemma:drift_condition} imply that the chain is non-evanescent \cite[Th.~9.4.1]{meyn2012markov}, and thus Harris recurrent \cite[Th.~9.2.2]{meyn2012markov}.
Since the recurrent chain admits a unique invariant probability measure $\pi_{\hat\chi}$ \cite[Th.~10.0.1]{meyn2012markov}, the $\psi$-irreducible chain $\{\hat{\chi}_\ell\}_{\ell \in \Z_{\geq 0}}$ is positive, and hence positive Harris recurrent \cite[Sec.~10.1]{meyn2012markov}.
Finally, for $\hat\chi_\ell \in \R^{n_x} \setminus \Dset$, we have
\begin{equation}
  \E\left[V_1(\hat{\chi}_{\ell+1}) \midd \hat{\chi}_\ell\right] - V_1(\hat{\chi}_\ell)
  \leq
  -f(\hat\chi_\ell) - 1
  =
  -\tilde{f}(\hat\chi_\ell).
\end{equation}
On the other hand, for any $\hat\chi_\ell \in \Dset$, using
\begin{align}
  &\E\left[V_1(\hat{\chi}_{\ell+1}) \midd \hat{\chi}_\ell\right] - V_1(\hat{\chi}_\ell) \\
  &=
  \beta^{(1)} + \gamma^{(1)} - f(\hat\chi_\ell) - \E\left[\bar g(\chi_\ell,\hat\chi_\ell,\bar w_\ell,\bar\omega_\ell)\midd \hat\chi_\ell\right]
\end{align}
and $\E[\bar g(\chi_\ell,\hat\chi_\ell,\bar w_\ell,\bar\omega_\ell)\, | \,\hat\chi_\ell]\ge 0$, we obtain
\begin{align}
  \E\left[V_1(\hat{\chi}_{\ell+1}) \midd \hat{\chi}_\ell\right] - V_1(\hat{\chi}_\ell) 
  &\leq \beta^{(1)} + \gamma^{(1)} - f(\hat\chi_\ell) \\
  &= -\tilde f(\hat\chi_\ell) + \beta^{(1)} + \gamma^{(1)} + 1.
\end{align}
Hence, with $b \coloneqq \beta^{(1)} + \gamma^{(1)} + 1$, it holds that
\begin{equation}
  \E\left[V_1(\hat{\chi}_{\ell+1}) \midd \hat{\chi}_\ell\right] - V_1(\hat{\chi}_\ell)
  \leq
  -\tilde{f}(\hat\chi_\ell) + b\mathbb{I}_{\Dset}(\hat\chi_\ell),
  \ \forall \hat\chi_\ell \in \R^{n_x},
\end{equation}
which establishes the $\tilde f$-ergodicity of the chain $\{\hat{\chi}_\ell\}_{\ell \in \Z_{\geq 0}}$ \cite[Th.~14.0.1]{meyn2012markov}.
\hfill\QED

\subsection{Proof of Theorem~\ref{thm:stability_guarantee}}
\label{append:proof_of_stability_guarantee}

Since $P^{(1)}_0$ is positive definite by Lemmas~\ref{lemma:positive_definiteness_Pp} and \ref{lemma:initial_value_P0}, from \cite[Cor.~10.4.2]{bernstein2018scalar}, there exist constants $0<\underline\lambda \leq \overline\lambda < \infty$ such that
\begin{equation}
  \underline\lambda \|x\|^2 \le V_1(x) \leq \overline\lambda \|x\|^2,\quad \forall x\in\R^{n_x}.
  \label{eq:norm-equiv}
\end{equation}
By Lemma~\ref{lemma:foster_lyapunov}, for the given initial condition $\hat{\chi}_0$, there exist a constant $c_0<\infty$ and $\ell_0\in\Z_{\geq 0}$ such that
\begin{equation}
  \sup_{\ell\geq \ell_0} \mathbb{E}\left[V_1(\chi_\ell) \midd \hat{\chi}_0\right] \leq c_0.
  \label{eq:V1-limsup}
\end{equation}
For the case where $0\le \ell<\ell_0$, since the plant is linear and the noise process has bounded covariance, we have $\mathbb{E}\left[\|\chi_\ell\|^2 \midd \hat{\chi}_0\right] < \infty$, and hence, by \eqref{eq:norm-equiv}, also $\mathbb{E}\left[V_1(\chi_\ell) \midd \hat{\chi}_0\right] < \infty$.
Therefore,
\begin{equation}
  \sup_{\ell\ge 0} \mathbb{E}\left[V_1(\chi_\ell) \midd \hat{\chi}_0\right] < \infty.
\end{equation}
Taking expectation and using the tower property yields
\begin{equation}
  \sup_{\ell\ge 0} \mathbb{E}\left[V_1(\chi_\ell)\right]
  =
  \sup_{\ell\ge 0}\mathbb{E}\left[\mathbb{E}\left[V_1(\chi_\ell) \midd \hat{\chi}_0\right]\right]
  < \infty.
  \label{eq:V1-unif}
\end{equation}
Combining \eqref{eq:norm-equiv} and \eqref{eq:V1-unif} yields
\begin{equation}
  \sup_{\ell\ge 0} \mathbb{E}\left[\|\chi_\ell\|^2\right]
  \leq \frac{1}{\underline\lambda} \sup_{\ell\ge 0}\mathbb{E}\left[V_1(\chi_\ell)\right]
  < \infty.
  \label{eq:chi2-unif}
\end{equation}

For a fixed $m \in \Mset $, we have
\begin{align}
  x_{k+1}
  &= A x_k + B u_k + w_k \\
  &= \Theta^{(m)}_{k-\ell h} x_k + \rho^{(m)}_{k-\ell h} B F^{(m)}_{k-\ell h} e_k + w_k,
  \label{eq:state-recursion}
\end{align}
for $k=\ell h,\ldots,(\ell+1)h-1$.
Defining
\begin{equation}
  \Xi^{(m)}_{(\tau:0)} \coloneqq \Theta^{(m)}_{\tau-1}\cdots\Theta^{(m)}_{0},\quad
  \Xi^{(m)}_{(0:0)} \coloneqq I_{n_x},
  \label{eq:Theta-Xi-def}
\end{equation}
and iterating \eqref{eq:state-recursion} from $k=\ell h$ to $k=\ell h+\tau$ with $\tau\in\{0,1,\dots,h-1\}$, we obtain
\begin{align}
  x_{\ell h+\tau}
  &= \Xi^{(m)}_{(\tau:0)} \chi_\ell \\
  &\quad + \sum_{i=0}^{\tau-1} \Xi^{(m)}_{(\tau:i+1)} \left(\rho^{(m)}_{i} B F^{(m)}_{i} e_{\ell h+i} + w_{\ell h+i}\right).
  \label{eq:unrolled}
\end{align}
Since $\Theta^{(m)}_{i}$ and $F^{(m)}_{i}$ are bounded for every $i \in \{0,1,\dots,h-1\}$ and $m \in \Mset$, we can define finite constants as
\begin{align}
  &\overline\Theta \coloneqq \max_{m,i} \|\Theta^{(m)}_{i}\|,\quad
  \overline F \coloneqq \max_{m,i} \|B F^{(m)}_{i}\|, \\
  &c_\Theta(\tau) \coloneqq \max_{m} \|\Xi^{(m)}_{(\tau:0)}\| \leq \overline\Theta^{\tau}.
  \label{eq:operator-bounds}
\end{align}
From the Cauchy--Schwarz inequality, which gives $(a+b+c)^2 \leq 3(a^2+b^2+c^2)$ for scalars $a$, $b$, and $c$, we obtain
\begin{align}
  &\mathbb{E}\left[\|x_{\ell h+\tau}\|^2\right] \\
  &\leq 3\mathbb{E}\left[\|\Xi^{(m)}_{(\tau:0)}\chi_\ell\|^2\right] \\
  &\quad + 3\sum_{i=0}^{\tau-1} \mathbb{E}\left[\|\Xi^{(m)}_{(\tau:i+1)} \rho^{(m)}_{i} B F^{(m)}_{i} e_{\ell h+i}\|^2\right] \\
  &\quad + 3\sum_{i=0}^{\tau-1} \mathbb{E}\left[\|\Xi^{(m)}_{(\tau:i+1)} w_{\ell h+i}\|^2\right] \\
  &\leq 3c_\Theta(\tau)^2\mathbb{E}\left[\|\chi_\ell\|^2\right] \\
  &\quad + 3\sum_{i=0}^{\tau-1} \overline\Theta^{2(\tau-i-1)} \overline F^{2} \tr(\Sigma) \\
  &\quad + 3\sum_{i=0}^{\tau-1} \overline\Theta^{2(\tau-i-1)} \tr\left(\Omega_w\right) \\
  &\leq a_\tau \mathbb{E}\left[\|\chi_\ell\|^2\right] + b_\tau,
  \label{eq:block-bound}
\end{align}
where $a_\tau$ and $b_\tau$ are defined as
\begin{align}
  a_\tau &\coloneqq 3 \overline\Theta^{2\tau}, \\
  b_\tau &\coloneqq 3\sum_{i=0}^{\tau-1} \overline\Theta^{2(\tau-i-1)} \left(\overline F^{2}\tr(\Sigma) + \tr(\Omega_w)\right).
\end{align}
Taking the supremum in $\ell$ in \eqref{eq:block-bound} and using \eqref{eq:chi2-unif} yields
\begin{equation}
  \sup_{\ell\ge0} \mathbb{E}\left[\|x_{\ell h+\tau}\|^2\right]
  \leq a_\tau \sup_{\ell\geq 0}\mathbb{E}\left[\|\chi_\ell\|^2\right] + b_\tau
  < \infty
\end{equation}
for $\tau=0,1,\dots,h-1$, which establishes \eqref{eq:mean_square_stability}.
\hfill\QED

\subsection{Proof of Theorem~\ref{thm:performance_guarantee}}
\label{append:proof_of_performance_guarantee}

To prove Theorem~\ref{thm:performance_guarantee}, we first rewrite the objective function $J^a$ under the policies $(\mu^{u, \text{ro}}, \mu^{\delta, \text{ro}})$.
For the policies $(\mu^{u, \text{ro}}, \mu^{\delta, \text{ro}})$, we have
\begin{align}
  &J^a(\mu^{u, \text{ro}}, \mu^{\delta, \text{ro}}) \\
  &= \limsup_{N \to \infty} \frac{1}{N} \E\left[ \sum_{k=0}^{N-1} x_k^\T Q x_k + u_k^\T R u_k + \theta \delta_k\right] \\
  &= \limsup_{L \to \infty} \frac{1}{hL} \E\Biggl[ \sum_{\ell=0}^{L-1} \sum_{\tau=0}^{h-1} x_{\ell h+\tau}^\T Q x_{\ell h+\tau} + u_{\ell h+\tau}^\T R u_{\ell h+\tau} \\
  &\qquad + \theta \delta_{\ell h+\tau}\Biggr] \\
  &= \limsup_{L \to \infty} \frac{1}{hL} \E\left[ \sum_{\ell=0}^{L-1} \bar{g}\left(\chi_\ell, \hat{\chi}_\ell, \bar{w}_\ell, \bar{\omega}_\ell\right) \right]. \label{eq:objective_function_rewrite}
\end{align}
The second equality in \eqref{eq:objective_function_rewrite} follows from Theorem~\ref{thm:stability_guarantee}, which admits
\begin{equation}
  \sup_{k\geq 0} \E\left[\|g(x_k, u_k, \delta_k)\|\right] < \infty.
\end{equation}

At iteration $\ell$, we have that
  \begin{align}
    &\E\Biggl[ \sum_{i=0}^{h-1} x_{\ell h+i}^\T Q x_{\ell h+i} + u_{\ell h + i}^\T R u_{\ell h+i} + \theta\delta_{\ell h+i} \\
    &\quad + x_{(\ell+1)h}\tilde{P}_px_{(\ell+1)h} \,\Big|\, \Iset_{\ell h} \Biggr] \\
    &= \E\Biggl[ \sum_{i=0}^{h-1} x_{\ell h+i}^\T Q x_{\ell h+i} + (\mu^{u,\text{ro}}_{\ell h+i}(\hat{\chi}_\ell))^\T R \mu^{u,\text{ro}}_{\ell h+i}(\hat{\chi}_\ell) \\
    &\quad + \theta \mu^{\delta,\text{ro}}_{\ell h+i}(\hat{\chi}_\ell) + x_{(\ell+1)h}\tilde{P}_px_{(\ell+1)h} \,\Big|\, \hat{\chi}_\ell \Biggr] \\
    &= \E\left[V_1(\chi_{\ell+1}) + \bar{g}\left(\chi_\ell, \hat{\chi}_\ell, \bar{w}_\ell, \bar{\omega}_\ell\right) \midd \hat{\chi}_\ell\right] \\
    &= \hat{\chi}_\ell P^{(\varphi(\hat{\chi}_\ell))}_0 \hat{\chi}_\ell + \tr\left(P^{(\varphi(\hat{\chi}_\ell))}_0 \Sigma\right) \\
    &\quad + \beta^{(\varphi(\hat{\chi}_\ell))} + \gamma^{(\varphi(\hat{\chi}_\ell))}.
  \end{align}
  From \eqref{eq:foster_lyapunov_function} and the fact that
  \begin{equation}
    \E\left[\chi_\ell^\T P^{(m)}_0 \chi_\ell \midd \hat{\chi}_\ell\right]  = \hat{\chi}_\ell^\T P^{(m)}_0 \hat{\chi}_\ell + \tr\left(P^{(m)}_0 \Sigma\right),
  \end{equation}
  holds for all $m \in \Mset$ \cite{astrom2006introduction}, it follows that
  \begin{align}
    &\E\left[V_1(\chi_{\ell+1}) + \bar{g}\left(\chi_\ell, \hat{\chi}_\ell, \bar{w}_\ell, \bar{\omega}_\ell\right) \midd \hat{\chi}_\ell\right] - \E\left[V_1(\chi_\ell) \midd \hat{\chi}_\ell\right] \\
    &= \hat{\chi}_\ell \left(P^{(\varphi(\hat{\chi}_\ell))}_0 - P^{(1)}_0 \right) \hat{\chi}_\ell + \tr\left(\left(P^{(\varphi(\hat{\chi}_\ell))}_0 - P^{(1)}_0\right)\Sigma\right) \\
    &\quad + \beta^{(\varphi(\hat{\chi}_\ell))} + \gamma^{(\varphi(\hat{\chi}_\ell))} \\
    &= \beta^{(1)} + \gamma^{(1)} - f(\hat{\chi}_\ell), \label{eq-append:h_step_gap}
  \end{align}
  where $\beta^{(1)}$ and $\gamma^{(1)}$ are given by \eqref{eq-append:beta1} and \eqref{eq-append:gamma1}, respectively, and $f$ is defined in \eqref{eq:def_fm}.
  Summing \eqref{eq-append:h_step_gap} over $\ell=0,1,\dots,L-1$ for $L \in \N$ yields
  \begin{align}
    &\sum_{\ell=0}^{L-1} \E\left[V_1(\chi_{\ell+1}) + \bar{g}\left(\chi_\ell, \hat{\chi}_\ell, \bar{w}_\ell, \bar{\omega}_\ell\right) \midd \hat{\chi}_\ell\right] \\
    &\quad - \sum_{\ell=0}^{L-1} \E\left[V_1(\chi_\ell) \midd \hat{\chi}_\ell\right] \\
    &= L\left(\beta^{(1)} + \gamma^{(1)}\right) - \sum_{\ell=0}^{L-1} f(\hat{\chi}_\ell). \label{eq-append:adding_gap}
  \end{align}
  Conditioning on $\hat{\chi}_0$ and applying both the tower property and the linearity of conditional expectation \cite[Th.~5.1.2]{durrett2010probability}, the left-hand side of \eqref{eq-append:adding_gap} becomes
  \begin{align}
    &\E\Biggl[ \sum_{\ell=0}^{L-1} \E\left[V_1(\chi_{\ell+1}) + \bar{g}\left(\chi_\ell, \hat{\chi}_\ell, \bar{w}_\ell, \bar{\omega}_\ell\right) \midd \hat{\chi}_\ell\right] \\
    &\quad - \sum_{\ell=0}^{L-1} \E\left[V_1(\chi_\ell) \midd \hat{\chi}_\ell\right] \,\Big|\, \hat{\chi}_0\Biggr] \\
    &= \E\left[\sum_{\ell=0}^{L-1} \bar{g}\left(\chi_\ell, \hat{\chi}_\ell, \bar{w}_\ell, \bar{\omega}_\ell\right) \midd \hat{\chi}_0\right] \\
    &\quad + \E\left[V_1(\chi_L) \midd \hat{\chi}_0\right] - \E\left[V_1(\chi_0) \midd \hat{\chi}_0\right].
  \end{align}
  Hence, we have
  \begin{align}
    &\E\left[\sum_{\ell=0}^{L-1} \bar{g}\left(\chi_\ell, \hat{\chi}_\ell, \bar{w}_\ell, \bar{\omega}_\ell\right) \midd \hat{\chi}_0\right] \\
    &= L\left(\beta^{(1)} + \gamma^{(1)}\right) - \E\left[\sum_{\ell=0}^{L-1} f(\hat{\chi}_\ell) \midd \hat{\chi}_0\right] \\
    &\quad - \E\left[V_1(\chi_L) \midd \hat{\chi}_0\right] + \E\left[V_1(\chi_0) \midd \hat{\chi}_0\right]. \label{eq:before_taking_limit}
  \end{align}
  Dividing both sides by $hL$ and taking the limit superior as $L \to \infty$, the left-hand side of \eqref{eq:before_taking_limit} is expressed as
  \begin{align}
    &\limsup_{L \to \infty} \frac{1}{hL} \E\left[\sum_{\ell=0}^{L-1} \bar{g}\left(\chi_\ell, \hat{\chi}_\ell, \bar{w}_\ell, \bar{\omega}_\ell\right) \midd \hat{\chi}_0\right] \\
    &= J^a(\mu^{u,\text{ro}}, \mu^{\delta,\text{ro}}). \label{eq:left}
  \end{align}
  On the right-hand side of \eqref{eq:before_taking_limit}, we obtain
  \begin{align}
    &\limsup_{L \to \infty} \frac{1}{h} \left(\beta^{(1)} + \gamma^{(1)}\right) \\
    &= \frac{1}{p} \left(\tr\left(\tilde{P}_p D_p \Omega_w^{(p)} D_p^\T\right) + \tr\left(M_p \Sigma\right) + d^a(p) + \theta\right) \\
    &= J^a(\mu^{u,\text{per}}, \mu^{\delta,\text{per}}), \label{eq:right1}
  \end{align}
  and
  \begin{equation}
    \limsup_{L \to \infty} \frac{1}{hL} \E\left[V_1(\chi_0) \midd \hat{\chi}_0\right] = 0. \label{eq:right2}
  \end{equation}
Combining \eqref{eq:left}, \eqref{eq:right1}, and \eqref{eq:right2} gives
\begin{align}
  &J^a(\mu^{u,\text{ro}}, \mu^{\delta,\text{ro}}) \leq J^a(\mu^{u,\text{per}}, \mu^{\delta,\text{per}}) \\
  &\qquad - \liminf_{L \to \infty} \frac{1}{hL }\E\left[V_1(\chi_L) \midd \hat{\chi}_0\right] \\
  &\qquad - \liminf_{L \to \infty} \frac{1}{hL} \E\left[\sum_{\ell=0}^{L-1} f(\hat{\chi}_\ell) \midd \hat{\chi}_0\right], \label{eq-append:main1}
\end{align}
where we use the subadditivity of the limit superior and the identity
$\limsup_{L\to\infty}(-a_L) = -\liminf_{L\to\infty} a_L$.

Finally, we show that the limit inferiors in \eqref{eq-append:main1} converge to nonnegative points.
From Lemma~\ref{lemma:foster_lyapunov}, it follows that
\begin{equation}
  \liminf_{L \to \infty} \frac{1}{hL} \E\left[V_1(\chi_L) \midd \hat{\chi}_0\right] = 0. \label{eq-append:main2}
\end{equation}
Moreover, for every initial condition $\hat{\chi}_0$, Theorem~\ref{thm:ergodicity} ensures
\begin{equation}
  \lim_{L \to \infty} \frac{1}{L}\E\left[\sum_{\ell=0}^{L-1} \tilde{f}(\hat{\chi}_\ell) \midd \hat{\chi}_0\right]
  =
  \int_{\R^{n_x}} \tilde{f}(\chi) \pi_{\hat\chi}(d\chi). \label{eq-append:ergodicity}
\end{equation}
The above integrand is strictly positive due to the open set irreducibility of $\{\hat{\chi}_\ell\}_{\ell \in \Z_{\geq 0}}$.
Using $\tilde{f}=f+1$ and \eqref{eq-append:ergodicity} yields
\begin{equation}
  \liminf_{L \to \infty} \frac{1}{hL} \E\left[\sum_{\ell=0}^{L-1} f(\hat{\chi}_\ell) \midd \hat{\chi}_0\right]
  =
  \tilde{c} - \frac{1}{h}, \label{eq-append:main3}
\end{equation}
where
\begin{equation}
  \tilde{c} \coloneqq \frac{1}{h} \int_{\R^{n_x}} \tilde{f}(\chi) \pi_{\hat\chi}(d\chi).
\end{equation}
Combining \eqref{eq-append:main1}, \eqref{eq-append:main2}, and \eqref{eq-append:main3} establishes \eqref{eq:performance_guarantee}.
\hfill\QED

\balance
\bibliographystyle{IEEEtran}
\bibliography{myrefs}

@article{antunes2014rollout,
  title = {Rollout event-triggered control: Beyond periodic control performance},
  author = {Antunes, D and Heemels, W P M H},
  journal = {IEEE Trans. Autom. Control},
  volume = {59},
  number = {12},
  pages={3296--3311},
  year = {2014}
}

@inproceedings{bommannavar2008optimal,
  title={Optimal control with limited control actions and lossy transmissions},
  author={Bommannavar, Praveen and Ba\c{s}ar, Tamer},
  booktitle={Proc. 47th IEEE Conf. Decis. Control},
  pages={2032--2037},
  year={2008}
}

@article{chan2007state,
  title={The state of the art of electric, hybrid, and fuel cell vehicles},
  author={Chan, Ching Chuen},
  journal={Proc. IEEE},
  volume={95},
  number={4},
  pages={704--718},
  year={2007},
  publisher={IEEE}
}

@inproceedings{cogill2009event,
  title={Event-based control using quadratic approximate value functions},
  author={Cogill, Randy},
  booktitle={Proc. 48h IEEE Conf. Decis. Control held jointly with 28th Chinese Control Conf.},
  pages={5883--5888},
  year={2009}
}

@article{demirel2016trade,
  title={On the trade-off between communication and control cost in event-triggered dead-beat control},
  author={Demirel, Burak and Gupta, Vijay and Quevedo, Daniel E and Johansson, Mikael},
  journal={IEEE Trans. Autom. Control},
  volume={62},
  number={6},
  pages={2973--2980},
  year={2016},
  publisher={IEEE}
}

@article{gao2011cardinality,
  title={Cardinality constrained linear-quadratic optimal control},
  author={Gao, Jianjun and Li, Duan},
  journal={IEEE Trans. Autom. Control},
  volume={56},
  number={8},
  pages={1936--1941},
  year={2011},
  publisher={IEEE}
}

@misc{grant2014cvx,
  title={{CVX}: Matlab software for disciplined convex programming},
  author={Grant, Michael and Boyd, Stephen},
  howpublished={\url{https://cvxr.com/cvx/}},
  year={2014}
}

@inproceedings{imer2006optimal,
  title={Optimal control with limited controls},
  author={Imer, Orhan C and Ba\c{s}ar, Tamer},
  booktitle={Proc. Amer. Control Conf.},
  pages={298--303},
  year={2006}
}

@article{joseph2020controllability,
  title={Controllability of linear dynamical systems under input sparsity constraints},
  author={Joseph, Geethu and Murthy, Chandra R},
  journal={IEEE Trans. Autom. Control},
  volume={66},
  number={2},
  pages={924--931},
  year={2020},
  publisher={IEEE}
}

@article{jovanovic2016controller,
  title={Controller architectures: Tradeoffs between performance and structure},
  author={Jovanovi{\'c}, Mihailo R and Dhingra, Neil K},
  journal={Eur. J. Control},
  volume={30},
  pages={76--91},
  year={2016},
  publisher={Elsevier}
}

@article{lin2013design,
  title={Design of optimal sparse feedback gains via the alternating direction method of multipliers},
  author={Lin, Fu and Fardad, Makan and Jovanovi{\'c}, Mihailo R},
  journal={IEEE Trans. Autom. Control},
  volume={58},
  number={9},
  pages={2426--2431},
  year={2013},
  publisher={IEEE}
}

@article{liu2003energy,
  title={Energy-efficient operation of rail vehicles},
  author={Liu, Rongfang Rachel and Golovitcher, Iakov M},
  journal={Transp. Res. Part A Policy Pract.},
  volume={37},
  number={10},
  pages={917--932},
  year={2003},
  publisher={Elsevier}
}

@article{nagahara2016maximum,
  title={Maximum hands-off control: A paradigm of control effort minimization},
  author={Nagahara, Masaaki and Quevedo, Daniel E and Ne{\v{s}}i{\'c}, Dragan},
  journal={IEEE Trans. Autom. Control},
  volume={61},
  number={3},
  pages={735--747},
  year={2016},
  publisher={IEEE}
}

@article{nagahara2024survey,
  title={A survey on compressed sensing approach to systems and control},
  author={Nagahara, Masaaki and Yamamoto, Yutaka},
  journal={Math. Control Signals Syst.},
  volume={36},
  number={1},
  pages={1--20},
  year={2024},
  publisher={Springer}
}

@article{nishida2024feedback,
  title={Feedback control balancing quadratic performance and input sparsity},
  author={Nishida, Shumpei and Okano, Kunihisa},
  journal={IEEE Control Syst. Lett.},
  volume={8},
  pages={970--975},
  year={2024},
  publisher={IEEE}
}

@article{pasand2018controllability,
  title={Controllability and stabilizability of multi-rate sampled data systems},
  author={Pasand, Mohammad Mahdi Share and Montazeri, Mohsen},
  journal={Syst. Control Lett.},
  volume={113},
  pages={27--30},
  year={2018},
  publisher={Elsevier}
}

@article{shi2013finite,
  title={Finite horizon {LQR} control with limited controller-system communication},
  author={Shi, Ling and Yuan, Ye and Chen, Jiming},
  journal={IEEE Trans. Autom. Control},
  volume={58},
  number={7},
  pages={1835--1841},
  year={2013},
  publisher={IEEE}
}

@article{sriram2022stabilizability,
  title={Stabilizability of linear dynamical systems using sparse control inputs},
  author={Sriram, Chandrasekhar and Joseph, Geethu and Murthy, Chandra R},
  journal={IEEE Trans. Autom. Control},
  volume={68},
  number={8},
  pages={5014--5021},
  year={2022},
  publisher={IEEE}
}

@article{eldar2010block,
  title={Block-sparse signals: Uncertainty relations and efficient recovery},
  author={Eldar, Yonina C and Kuppinger, Patrick and Bolcskei, Helmut},
  journal={IEEE Trans. Signal Process.},
  volume={58},
  number={6},
  pages={3042--3054},
  year={2010},
  publisher={IEEE}
}

@book{astrom2006introduction,
  title={Introduction to Stochastic Control Theory},
  author={{\AA}str{\"o}m, Karl J},
  year={2006},
  publisher={Dover Publications}
}

@book{bernstein2018scalar,
  title={Scalar, Vector, and Matrix Mathematics: Theory, Facts, and Formulas},
  author={Bernstein, Dennis S},
  year={2018},
  publisher={Princeton University Press}
}

@book{bertsekas2017dynamic-i,
  title={Dynamic Programming and Optimal Control: Volume I},
  author={Bertsekas, Dimitri P},
  edition={4},
  year={2017},
  publisher={Athena Scientific}
}

@book{bertsekas1976dynamic,
  title={Dynamic Programming and Stochastic Control},
  author={Bertsekas, Dimitri P},
  year={1976},
  publisher={Academic Press}
}

@book{bertsekas2012dynamic-ii,
  title={Dynamic Programming and Optimal Control: Volume II},
  author={Bertsekas, Dimitri P},
  edition={4},
  year={2012},
  publisher={Athena Scientific}
}

@book{durrett2010probability,
  title={Probability: Theory and Examples},
  author={Durrett, Rick},
  edition={4},
  year={2010},
  publisher={Cambridge University Press}
}

@book{meyn2012markov,
  title={Markov Chains and Stochastic Stability},
  author={Meyn, Sean P and Tweedie, Richard L},
  edition={2},
  year={2012},
  publisher={Cambridge University Press}
}

@book{rudin1976principles,
  title={Principles of Mathematical Analysis},
  author={Rudin, Walter},
  edition={3},
  year={1976},
  publisher={McGraw-Hill}
}

@book{rudin1987real,
  title={Real and Complex Analysis},
  author={Rudin, Walter},
  edition = {3},
  year={1987},
  publisher={McGraw-Hill}
}

\end{document}